\documentclass[useAMS,usenatbib]{mnras}

\usepackage{url}
\usepackage[section]{placeins}
\usepackage{graphics,graphicx,times}
\usepackage[caption=false]{subfig}
\usepackage{pdflscape}
\usepackage{amsmath} 
\usepackage{amssymb}
\usepackage{deluxetable,natbib}

\newcommand{\nii}{[N \textsc{ii}]}
\newcommand{\sii}{[S \textsc{ii}]}
\newcommand{\oi}{[O \textsc{i}]}
\newcommand{\oii}{[O \textsc{ii}]}
\newcommand{\oiii}{[O \textsc{iii}]}
\newcommand{\ha}{H$\alpha$}
\newcommand{\hb}{H$\beta$}

\def\ltsima{$\; \buildrel < \over \sim \;$}
\def\simlt{\lower.5ex\hbox{\ltsima}}
\def\gtsima{$\; \buildrel > \over \sim \;$}
\def\simgt{\lower.5ex\hbox{\gtsima}}

\title[Spatially Resolving the SFMS]{The SAMI Galaxy Survey: Spatially Resolving the Main Sequence of Star Formation} 

\author[Medling et al.]{Anne M. Medling$^{1,2}$\thanks{Hubble Fellow; amedling@caltech.edu}, 
Luca Cortese$^3$, 
Scott M. Croom$^{4,5}$, 
Andrew W. Green$^6$,
Brent Groves$^2$,
\newauthor 
Elise Hampton$^2$,
I-Ting Ho$^7$, 
Luke J. M. Davies$^3$,
Lisa J. Kewley$^2$, 
Amanda J. Moffett$^8$,
\newauthor
Adam L. Schaefer$^4$, 
Edward Taylor$^9$, 
Tayyaba Zafar$^6$,
Kenji Bekki$^3$,
Joss Bland-Hawthorn$^4$,
\newauthor
Jessica V. Bloom$^4$,
Sarah Brough$^{6,10,5}$,
Julia J. Bryant$^{4,6,5}$,
Barbara Catinella$^3$,
Gerald Cecil$^{11}$,
\newauthor
Matthew Colless$^2$,
Warrick J. Couch$^6$,
Michael J. Drinkwater$^{12,5}$,
Simon P. Driver$^3$,
\newauthor
Christoph Federrath$^2$,
Caroline Foster$^6$,
Gregory Goldstein$^{13}$,
Michael Goodwin$^6$,
\newauthor
Andrew Hopkins$^6$,
J. S. Lawrence$^6$,
Sarah K. Leslie$^{7}$,
Geraint F. Lewis$^4$,
Nuria P. F. Lorente$^6$,
\newauthor
Matt S. Owers$^{13,6}$,
Richard McDermid$^{13,6}$,
Samuel N. Richards$^{14}$,
Robert Sharp$^2$,
\newauthor
Nicholas Scott$^4$,
Sarah M. Sweet$^{9}$,
Dan S. Taranu$^{3,5}$,
Edoardo Tescari$^{15,5}$,
Chiara Tonini$^{15}$,
\newauthor
Jesse van de Sande$^4$, 
C. Jakob Walcher$^{16}$,
and Angus Wright$^{17}$
\\
$^1$Cahill Center for Astronomy and Astrophysics, California Institute of Technology, MS 249-17, Pasadena, CA 91125, USA \\
$^2$Research School of Astronomy \& Astrophysics, Australian National University, Canberra, ACT 2611, Australia \\
$^3$International Centre for Radio Astronomy Research, University of Western Australia, 35 Stirling Highway, Crawley, WA 6009, Australia \\
$^4$Sydney Institute for Astronomy, School of Physics, University of Sydney, NSW 2006, Australia \\
$^5$ARC Centre of Excellence for All-sky Astrophysics (CAASTRO) \\
$^6$Australian Astronomical Observatory, PO Box 915, North Ryde, NSW 1670, Australia \\
$^7$Max-Planck-Institut f{\"u}r Astronomie, K{\"o}nigstuhl 17, 69117 Heidelberg, Germany \\
$^8$Department of Physics \& Astronomy, Vanderbilt University, 2401 Vanderbilt Place, Nashville, TN 37240, USA\\
$^9$Centre for Astrophysics and Supercomputing, Swinburne University of Technology, PO Box 218, Hawthorn, VIC 3122, Australia \\
$^{10}$School of Physics, University of New South Wales, NSW 2052, Australia \\
$^{11}$Department of Physics and Astronomy, University of North Carolina, Chapel Hill, NC 27599, USA \\
$^{12}$School of Mathematics and Physics, University of Queensland, QLD 4072, Australia \\
$^{13}$Department of Physics and Astronomy, Macquarie University, NSW 2109, Australia \\
$^{14}$SOFIA Operations Center, USRA, NASA Armstrong Flight Research Center, 2825 East Avenue P, Palmdale, CA 93550, USA \\
$^{15}$School of Physics, The University of Melbourne, Parkville, VIC 3010, Australia \\
$^{16}$Leibniz-Institut f{\"u}r Astrophysik Potsdam (AIP), An der Sternwarte 16 D-14482, Potsdam, Germany  \\
$^{17}$Argelander-Institut f{\"u}r Astronomie (AIfA), Universit{\"a}t Bonn, Auf dem H{\"u}gel 71, D-53121 Bonn, Germany
}

\begin{document}
\date{Accepted 2018 January 10; Received 2018 January 10; In original form 2017 May 19}
\label{firstpage}
\maketitle

\begin{abstract}
We present the $\sim$800 star formation rate maps for the SAMI Galaxy Survey based on \ha~emission maps, corrected for dust attenuation via the Balmer decrement, that are included in the SAMI Public Data Release 1.  We mask out spaxels contaminated by non-stellar emission using the \oiii/\hb, \nii/\ha, \sii/\ha, and \oi/\ha~line ratios.
Using these maps, we examine the global and resolved star-forming main sequences of SAMI galaxies as a function of morphology, environmental density, and stellar mass.  Galaxies further below the star-forming main sequence are more likely to have flatter star formation profiles.
Early-type galaxies split into two populations with similar stellar masses and central stellar mass surface densities.  The main sequence population has centrally-concentrated star formation similar to late-type galaxies, while galaxies $>$3$\sigma$ below the main sequence show significantly reduced star formation most strikingly in the nuclear regions.  The split populations support a two-step quenching mechanism, wherein halo mass first cuts off the gas supply and remaining gas continues to form stars until the local stellar mass surface density can stabilize the reduced remaining fuel against further star formation.
Across all morphologies, galaxies in denser environments show a decreased specific star formation rate from the outside in, supporting an environmental cause for quenching, such as ram-pressure stripping or galaxy interactions.  
\end{abstract}

\begin{keywords}
Surveys -- galaxies: star formation -- galaxies: evolution
\end{keywords}

\section{Introduction: The Star-Forming Main Sequence}

The stellar masses (M$_{*}$) and star formation rates (SFRs) of star-forming galaxies follow a tight relation called the star-forming main sequence \citep[SFMS;][]{Brinchmann04, Daddi07, Elbaz07, Noeske07a,Noeske07b}.  This relation takes the form SFR $\propto$ M$_{*}^{\beta}$, with most studies finding $\beta\sim$0.7-1.0 over the stellar mass range 10$^{7}$ - 10$^{10}$ M$_{\sun}$ \citep[e.g.][]{Santini09, Speagle14, Lee15}.  This tight relation \citep[scatter $\sim$0.2-0.35 dex;][]{Speagle14} indicates a nearly constant specific star formation rate (sSFR $\equiv$ SFR/M$_{*}$) in a given redshift bin; this typical sSFR increases with redshift \citep{Daddi07,Elbaz07,Whitaker12,Speagle14}.  Starburst galaxies lying above the main sequence may have an enhanced star formation efficiency (star formation rate compared to the gas mass; SFR/M$_{gas}$) rather than increased gas fractions \citep{Daddi10b, Tacconi10, Tacconi13, Silverman15, Scoville16}.  Galaxies falling below the main sequence have had their star formation suppressed or quenched.

Galaxies stop producing stars when cold gas is no longer able to feed star formation.
A variety of mechanisms can cause this quenching: 
gas could be removed from galaxies through outflows or tidal stripping, or gas has stopped accreting and the remaining gas has been consumed by star formation, or the gas that does remain in the galaxy is sufficiently heated or turbulent to be stable against gravitational collapse \citep{Federrath17, Zhou17}.
The study of quenching mechanisms has been fraught with the common astronomy practice of phenomenological naming schemes.  We attempt to explain some of the most common here:
\begin{itemize}
\item ``Morphological quenching'': coined by \citet{Martig09} to describe the stabilizing effects galactic bulges can have on galaxy discs.  In this framework, the stellar potential of the bulge increases the Toomre Q parameter \citep{Toomre64} by increasing the orbital velocities such that a gaseous disc is no longer self-gravitating, perhaps related to velocity shearing due to the steep potential wells in the nuclei of bulges \citep{Federrath16,Krumholz17}.
Additionally, stars in a spheroid instead of a stellar disc would fail to contribute to the gravitational collapse of the gas disc.  
\item ``Mass quenching": generic term used to describe quenching processes that scale with galaxy mass \citep{Peng10}.  Halo quenching is one of these processes.
\item ``Environment(al) quenching": generic term used to describe quenching processes that scale with the cosmological environmental densities \citep{Peng10,Peng12}.  Strangulation, ram-pressure stripping, and halo quenching can all be mechanisms that act in environmental quenching.
\item ``Halo quenching": as gas accretes onto a galaxy with a dark matter halo more massive than $\sim$10$^{12}$M$_{\sun}$, it forms a virial shock where the gas travels faster than the speed of sound \citep{Birnboim03,Keres05,Dekel06,Woo13}.  The shocked gas is heated, preventing additional cold gas accretion to the galaxy disc and thus limiting further star formation fuel.  Halo quenching might occur in a massive galaxy with a massive halo, or in a satellite galaxy embedded in a more massive halo.
\item ``Strangulation": used to describe when a galaxy stops replenishing its star-forming fuel \citep[e.g.][]{Larson80,Balogh00a,Balogh00b,Peng15}, e.g. through halo quenching.
\item ``Ram-pressure stripping": when galaxies fall into a cluster, their halos can be stripped off while moving through the hot intracluster medium \citep{Gunn72}.
\item ``Inside-out quenching": generic term where nuclear star formation is shut off through any mechanism before star formation at larger radii ceases \citep[e.g.][]{Tacchella15}.  We note that ``inside-out quenching'' here is related to but distinct from the ``inside-out growth'' of disks \citep{White91, Mo98}, which describes the overall pattern of stellar mass buildup rather than the cessation of star formation \citep[e.g.][]{MunozMateos07,Perez13,IbarraMedel16,Goddard17a,Goddard17b}.
\item ``Stellar Feedback'': the process through which ongoing star formation affects the host galaxy, possibly inhibiting future star formation.  Stellar feedback includes winds from evolved stars and the energy injected into the interstellar medium from supernovae \citep[e.g.][]{Strickland02,Federrath15}. 
\item ``AGN Feedback": the process through which an active galactic nucleus affects the host galaxy, possibly inhibiting star formation.  See e.g. \citet{Fabian12} for a review.
\end{itemize}

Observational studies continue to build up the statistical properties of quenched and star-forming galaxy populations to find clues of physical processes that may be driving this transformation.  \citet{Tully82} first noted a morphological split in colour-magnitude space: late-type galaxies following a blue, gas-rich, star-forming sequence before transitioning quickly to the red, quenched, early-type galaxy sequence; this bimodal distribution is more evident in recent larger surveys (\citealt{Baldry04, Brammer09}; but see also \citealt{Feldmann17}).  The central stellar mass surface density may be the strongest predictor of the quenched population \citep[e.g.][]{Bezanson09, Whitaker17}, and recent simulations point to the central gas density as another relevant quantity \citep{Tacchella16a}.
A possible flattening of the star-forming main sequence slope at higher stellar masses (\citealt{Karim11,Bauer13,Whitaker14,Lee15}; but see \citealt{RenziniPeng15}) suggests that stellar mass may play a strong role in the quenching of galaxies, but the mechanism through which this occurs is not yet clear.  

The star-forming main sequence has traditionally included SFR and M$_{*}$ values integrated over an entire galaxy.  
\citet{Wuyts13} used HST grism spectroscopy and multiband imaging to first resolve the star-forming main sequence into spatial regions on $\sim$1 kpc scales for galaxies at $0.7<z<1.5$.  
Large integral field spectroscopy surveys of local galaxies are now also producing spatially resolved star formation maps and stellar mass maps to determine how sSFR varies in a single galaxy.  
One such study is the Calar Alto Legacy Integral Field Area survey \citep[CALIFA;][]{CALIFA, CALIFA_DR2}, which contains optical integral field spectroscopy for $\sim$600 local galaxies of a range of Hubble types, masses, luminosities, and colours.  The CALIFA galaxies show a spatially resolved main sequence of star formation regardless of dominant ionization source of the host galaxy or its integrated stellar mass \citep{Sanchez13,Cano-Diaz16}.  
\citet{GonzalezDelgado16} split the CALIFA sample into Hubble types and found that the sSFR has a radial dependence that varies by morphological type.  Early-type galaxies have sSFR $\sim$2 orders of magnitude lower in the nuclei than in their outskirts; this gradient flattens for progressively later-type galaxies.  This difference is interpreted as inside-out quenching. 

Using integral field spectroscopy of 500+ galaxies from the Mapping Nearby Galaxies at the Apache Point Observatory \citep[MaNGA;][]{MaNGA} survey, \citet{Belfiore17a} found further evidence of inside-out quenching.  
Galaxies with emission lines dominated by post-AGB stars (i.e. older stellar populations) in the centres only fall $\sim$1dex below the star-forming main sequence.  Galaxies that show this quenching signature across the entire galaxy, on the other hand, lie completely off the main sequence, showing overall SFRs similar to quiescent galaxies.  
Further, \citet{Belfiore17b} found that green valley galaxies have reduced specific star formation rates at all radii compared to blue cloud galaxies at the same stellar mass.
A study of twelve prototype-MaNGA galaxies also showed that galaxies with quiescent nuclei are likely to have positive radial gradients in specific star formation rates (via proxies like the equivalent width of \ha), whereas galaxies with star-forming nuclei show flat profiles \citep{Li15}.

\citet{Schaefer16} used similar data from the Sydney-AAO Multi-Object Integral field spectrograph (SAMI) Galaxy Survey \citep{Allen15, Green17} to approach this question from an environmental perspective.  They found that galaxies in denser environments show steeper SFR gradients and lower integrated SFRs.  These results are demonstrative of the environmental aspect of quenching, wherein close interactions and/or pressure from the intragroup/intracluster medium can strip or heat gas in the outskirts of a galaxy, causing star formation to quench from the outside-in rather than the inside-out \citep{Peng12,Peng15}.

In this paper, we use data from the SAMI Galaxy Survey \citep{Bryant15} to examine the spatially resolved star-forming main sequence. 
In Section~\ref{data}, we discuss the SAMI Galaxy Survey and our selection criteria.  We place these galaxies on the star-forming main sequence using integrated SFRs and total stellar masses in Section~\ref{global} and using spatially resolved SFRs from \ha~emission maps and stellar mass maps in Section~\ref{local}.  We discuss quenching mechanisms in Section~\ref{discussion} and present our conclusions in Section~\ref{summary}.  

\section{Data from the SAMI Galaxy Survey}
\label{data}

The SAMI Galaxy Survey \citep{Bryant15} uses the Anglo-Australian Telescope and will contain $\sim$3600 galaxies at z$<$0.1 covering a range of stellar masses ($10^{8}-10^{11.5}$ M$_{\sun}$) and environments drawn from the Galaxy And Mass Assembly survey \citep[GAMA;][]{GAMA} with additional galaxies from eight rich clusters with virial masses up to $1.5\times10^{15}$ M$_{\sun}$ \citep{Owers17}.  With the SAMI instrument \citep{Croom12}, each galaxy is centred on a 61-fibre hexabundle \citep{BlandHawthorn11,Bryant11,Bryant14}, which feeds the spatially-resolved spectra to the AAOmega spectrograph \citep{Sharp06}.  Using up to seven pointings per galaxy, the fibre spectra are weighted and combined to produce red (4700-6300\AA, R$\sim$4500) and blue (3700-5800\AA, R$\sim$1700) data cubes with 0\farcs5$\times$0\farcs5 spaxels \citep{Allen15, Sharp15}.  This paper uses the SAMI internal DR0.9 data, which contains the first 1296 galaxies; $\sim$800 of these are publicly released in DR1 \citep{Green17}.  

\subsection{Value-Added Data Products}
\begin{figure*}
\centering
\includegraphics[width=6inches,trim=0.8cm 0.3cm 0cm 0cm]{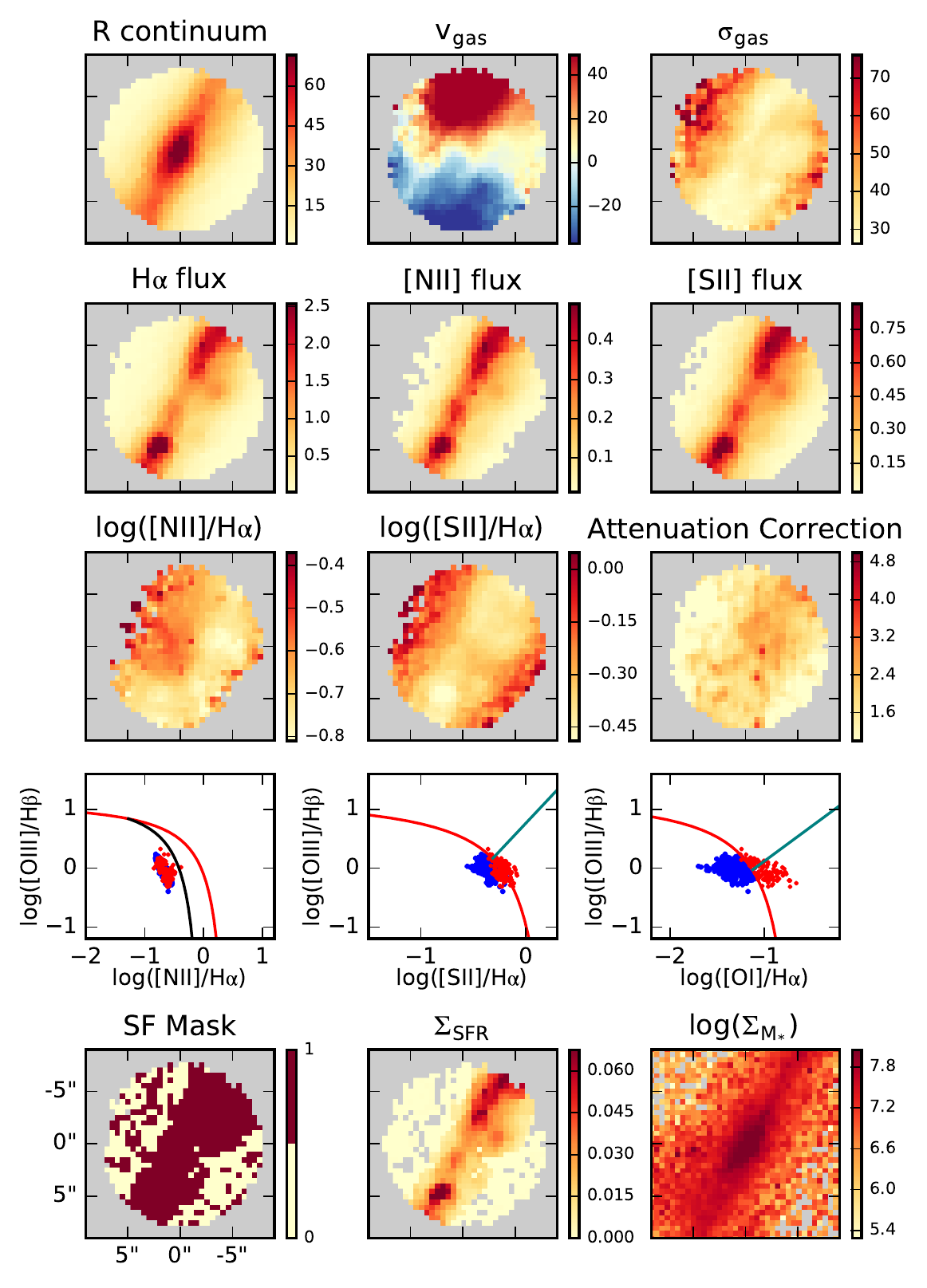}
\caption{Derived quantities from SAMI data for the galaxy GAMA 31452.  Panels show in the top row, right to left: maps of SAMI red arm continuum emission (in 10$^{-16}$ erg s$^{-1}$ cm$^{-2}$ \AA$^{-1}$ spaxel$^{-1}$), ionized gas velocity and ionized gas velocity dispersion (in km s$^{-1}$); second row: \ha~emission flux, \nii~emission flux, and \sii~emission flux (in 10$^{-16}$ erg s$^{-1}$ cm$^{-2}$ spaxel$^{-1}$); third row: log(\nii/\ha), log(\sii/\ha), attenuation correction map from Section~\ref{extinctionmaps}; fourth row: \nii, \sii~and \oi~diagrams with diagnostic lines from \citet{Kewley06}; fifth row: star formation mask from Section~\ref{SFmasks}, star formation rate surface density (in M$_{\sun}$ yr$^{-1}$ kpc$^{-2}$) from Section~\ref{SFRmaps}, and log of the stellar mass surface density (in M$_{\sun}$ kpc$^{-2}$) from Section~\ref{stellarmassmaps}.
} 
\label{dataproductexample}
\end{figure*}

Our SAMI star formation rates are based on \ha~emission.  The radiation fields from O and B stars in HII regions ionize the surrounding gas, making this emission line a good tracer for star formation on $\sim$10 Myr timescales \citep[e.g.][]{Kennicutt94}.  In the following subsections, we describe our method for deriving the star formation rate based on the \ha~and other emission line fluxes.  We note that the released Balmer emission line fluxes in SAMI have errors that are corrected to include stellar absorption uncertainties.  The current versions of the emission line fits (Version 03), attenuation correction maps (V05), star formation masks (V04), and star formation rate maps (V05) described below are included in the first SAMI Public Data Release \citep{Green17}, and made available at \url{datacentral.aao.gov.au}.  An example of each data product is shown for the galaxy GAMA 31452 in Figure~\ref{dataproductexample}.

\subsubsection{Emission Line Fits}

Each data cube was processed with \texttt{LZIFU} to subtract the stellar continuum and fit the emission lines \citep{LZIFU}.  This routine simultaneously fits \ha, \hb, \nii, \sii, \oi, \oii, and \oiii~with up to three Gaussian components each.  Full details on the implementation of \texttt{LZIFU} to SAMI data can be found in \citet{Green17}.

We determined the number of components appropriate for each spaxel using an artificial neural net \citep[\texttt{LZComp};][]{Hampton17}, trained by five SAMI astronomers.  We adopted the results from \texttt{LZComp} with the additional stipulation that the \ha~emission line must have a signal-to-noise ratio of at least 5 in each component; if it did not, the number of components for that spaxel was reduced until the stipulation was met (or until only one component remained).  We note that in cases where multiple components are recommended, we use the star formation calculated from summing the emission line fluxes across all components.  By allowing multiple components instead of single-component fits, we improve the accuracy of our fits in cases of beam smearing (showing multiple kinematic components that are both star-forming) or other complex line profiles in high signal-to-noise ratio cases.

One key feature of SAMI emission line fits is the adjustment of the Balmer flux uncertainties to account for uncertainties in stellar absorption correction.  Because \ha~and \hb~emission lie on top of stellar absorption features, the measured fluxes of these lines depend heavily on what stellar populations are fit and removed before line-fitting takes place.  The absorption correction can vary significantly for populations of different ages and depending on what set of stellar templates are used \citep[e.g.][]{Groves12}.  This variation may systematically affect the Balmer flux measurements, and definitively should increase the uncertainty in the line measurements.  To account for this uncertainty, we include the error on the Balmer absorption correction.

Balmer absorption in stellar populations is strongly correlated with the 4000\AA~break. We therefore estimate our errors on the Balmer absorption of the continuum using the errors on the D$_{n}$4000 index \citep{Balogh99}, which we measured by J. Moustakas' IDL routine \texttt{spectral\_indices}. This error in the Balmer continuum absorption equivalent width ($\delta_{\text{H}x,contEW}$) is then proportional to the extra uncertainty translated to the emission line fluxes. The Balmer continuum absorption is spread out over a wide spectral range, so only a fraction of it (based on the emission line width and here denoted by the coefficient $A_{\text{H}x}$) is propagated to the emission line flux. We calculate the total Balmer flux uncertainty $\delta_{\text{H}x,tot}$ from the original emission-line-only fit uncertainty $\delta_{\text{H}x,emission}$ using the equations: \\

$\delta_{\text{H}x,tot}^{2} = \delta_{\text{H}x,emission}^{2} + (A_{\text{H}x} \delta_{\text{H}x,contEW} f_{cont})^2 $

$A_{\text{H}\alpha} = 0.0504933 + 0.00605358\sigma $

$A_{\text{H}\beta} = 0.0258323 + 0.00481798\sigma $\\

\noindent where $\sigma$ is the velocity dispersion of the emission line, and $f_{cont}$ is the continuum level around the emission line, used to convert the equivalent width error into flux units. 
The $A_{\text{H}x}$ equations were calculated using the MILES stellar template libraries \citep{Vazdekis10} at typical ranges of metallicity, stellar velocity dispersion, and stellar population age.

These corrected uncertainties are the uncertainties on the emission line fluxes for Balmer emission lines included in the public database, and should be used as one would normally use uncertainties.  

\subsubsection{Attenuation Correction Maps}
\label{extinctionmaps}

Star formation rates measured from \ha~emission maps must be corrected for extinction, which is commonly done using the Balmer decrement $\frac{f_{\text{H}\alpha}}{f_{\text{H}\beta}}$.  However, this ratio of lines is impacted by aliasing introduced by the SAMI observing process.  As described in detail in \citet{Green17}, our data reduction process does not completely remove the effects of aliasing due to differential atmospheric refraction (DAR).  This aliasing means that the PSFs of two widely-separated wavelengths (such as \ha~and \hb) can vary, resulting in excess noise in the ratio of these wavelengths (i.e. $\frac{f_{\text{H}\alpha}}{f_{\text{H}\beta}}$).  This excess noise can be seen when comparing the flux ratio in one spaxel to its neighbours: because our PSF is oversampled, you would expect the variation between two neighbouring spaxels to be distributed roughly normally, with a $\sigma$ similar to that calculated formally from the uncertainties in the individual flux maps.  

\begin{figure}
\includegraphics[scale=0.92,trim=0.8cm 0.3cm 0cm 0cm]{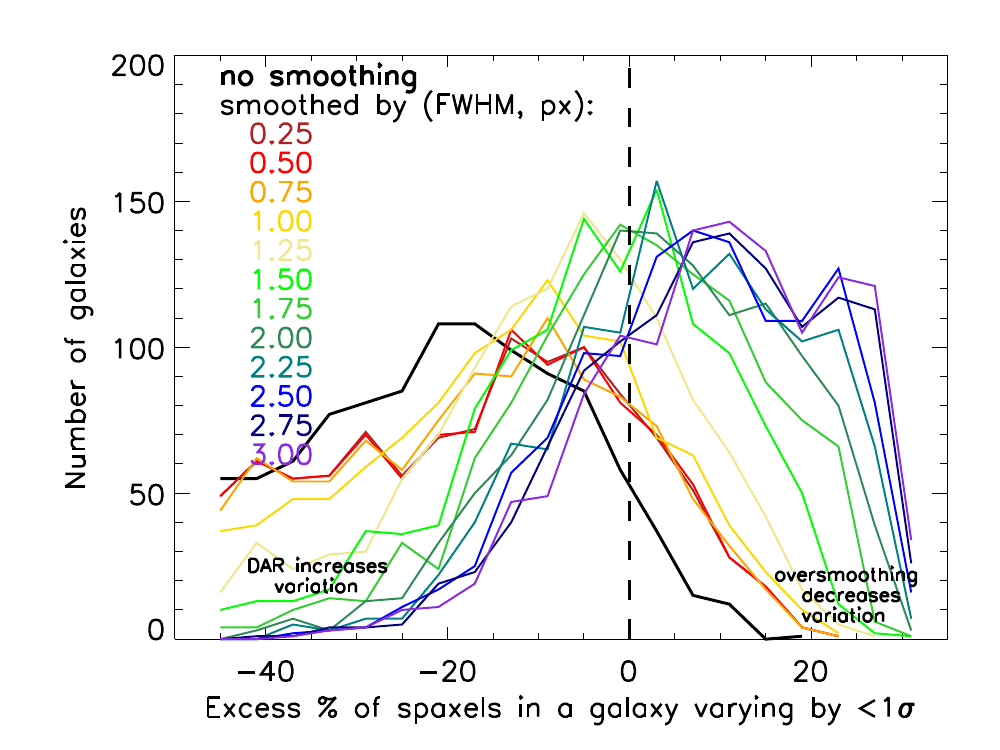}
\caption{Distributions of the excess percentages of spaxels in SAMI DR0.9 galaxies with Balmer decrement ($\frac{f_{\text{H}\alpha}}{f_{\text{H}\beta}}$) variations (median variation between four neighbouring spaxels) less than the formal uncertainty in the Balmer decrement.  Percentages are normalized such that 0 = 68\%, the percentage expected of a normal distribution with appropriate errors.  Aliasing due to differential atmospheric refraction (DAR) causes raw (unsmoothed, thick black line) ratios to show too few spaxels that vary smoothly given their uncertainties.  Spatially smoothing the line ratio maps by a Gaussian kernel of varying FWHM (colours) demonstrates that DAR artifacts can be compensated for and a typical amount of variation recovered.
} 
\label{DARcorrection}
\end{figure}

To demonstrate this, we determined the percentage of spaxels in each SAMI galaxy with the variation of the Balmer decrement (based on the median variation in four neighbouring spaxels) less than the expected 1$\sigma$ uncertainty based on the emission line flux uncertainties.  In Figure~\ref{DARcorrection}, we show the distribution of this percentage in SAMI galaxies (black line). 
We have subtracted 68\%, the percentage of spaxels expected if the uncertainties were normally distributed.
It is clear from this figure that the raw (unsmoothed) line ratio shows neighbouring spaxels vary far more than is predicted by their uncertainties.  

We spatially smooth the ratio map by a truncated 5x5 spaxel Gaussian kernel with a full-width at half maximum (FWHM) varying from 0.25-3.0 spaxels (coloured lines, Figure~\ref{DARcorrection}) to demonstrate that an appropriate set of smoothing compensates for the aliasing effect.  We find that variations are distributed roughly normally (centred about 0 in Figure~\ref{DARcorrection}) when smoothed with a kernel of FWHM 1.5-1.75 spaxels.  
We emphasize that this smoothing technique does not affect the overall measurement of the Balmer decrement, leaving the median level of dust across each galaxy unchanged.

We therefore extinction-correct our \ha~emission maps using the Balmer decrement smoothed by a 5x5 truncated Gaussian kernel with FWHM=1.6 spaxels ($(\frac{f_{\text{H}\alpha}}{f_{\text{H}\beta}})_{sm}$).  Note that we mask out spaxels with \ha~or \hb~signal-to-noise ratios SNR$<$3 before smoothing, to avoid spreading high-error regions.  Assuming Case B recombination \citep[following][]{Calzetti01}, the expected intrinsic value for the flux ratio $\frac{f_{\text{H}\alpha}}{f_{\text{H}\beta}}$ is 2.86; we obtain the attenuation correction factor $F_{attenuate}$ following the Cardelli extinction law \citep{Cardelli89}, which has a reddening slope of 2.36: \\

$F_{attenuate} =  (\frac{1}{2.86} (\frac{f_{\text{H}\alpha}}{f_{\text{H}\beta}})_{sm})^{2.36}$ \\

We set the correction factor $F_{attenuate} = 1$ and the associated error $\delta F_{attenuate} = 0$ for spaxels with no \hb~detection and for spaxels with a Balmer decrement less than 2.86.  
We note that spaxels might lack a detection in \hb~either because of high extinction or because of low SNR or data artifacts and for now avoid attenuation corrections based on limits.

\subsubsection{Star Formation Masks}
\label{SFmasks}

\ha~is not only emitted by the H{\sc ii} regions surrounding recent star-formation; it can also arise from gas photoionized by an active galactic nucleus (AGN) or collisionally ionized in interstellar shocks \citep{Dopita76, Shull79}.  These different ionization mechanisms lead to different conditions in the ionized gas, such as increased temperatures, resulting in different emission line ratios.  Key optical diagnostic line ratio diagrams plot \oiii/\hb~ versus \nii/\ha~\citep{BPT}, \sii/\ha, or \oi/\ha~\citep{VO87}; the position of a spectrum in these diagrams indicates whether the emission is dominated by star formation or substantially contaminated by another effect.

We classify each spaxel using the total emission line fluxes (i.e. not split between multiple components) using the diagnostic scheme of \citet{Kewley06}.  A spaxel is considered `star formation dominant' if all high signal-to-noise ratio (SNR$>$5) line ratios fall in the HII region portions of the emission line diagnostic diagrams.  That is, a strong \oi~detection is not required to classify a spectrum as star-forming; but if \oi~is detected, the spectrum must have an \oi/\ha~ratio consistent with photoionization by an HII region.  

The \citet{Kewley06} classification requires reliable \oiii/\hb~ratios, but for many spaxels of SAMI galaxies, the \oiii~line may not be detected.  In the case where the \oiii~emission has a SNR$<$5, we classify the spaxel as star-forming if the log(\nii/\ha)~ratio is less than -0.4.  
We also classify spaxels as star-forming if all forbidden lines have SNR less than 5 \textit{and} less than that of \ha.

Our conservative approach to SF classifications produces a clean but not complete sample of star-forming spaxels.  To avoid contamination, we intentionally remove star forming regions that are cospatial with winds or LINER/AGN emission that can significantly affect the line ratios and therefore the \ha~flux.  Young, low-metallicity star-forming regions with strongly-ionizing Wolf-Rayet stars, residual star formation near post-AGB stars, and diffuse ionized gas from HII regions may be missed.

We note that using line ratios alone can produce degeneracies between weak AGN photoionization, LINER-like ionization, and ionization in shock-heated gas.  For this reason, we only use this classification to produce masks indicating spaxels dominated by star formation (maps that are 1 for `star-forming' and 0 for `other').  Full classification separating star formation from AGN and shocks will be presented in a future paper.

\subsubsection{SFR Maps}
\label{SFRmaps}
We multiply the \ha~flux maps by the attenuation correction maps and star formation masks to convert the measured \ha~flux to the intrinsic \ha~flux from star formation.  We convert this intrinsic \ha~flux to luminosity using the distance to each galaxy, calculated using the flow-corrected redshifts \texttt{z\_tonry\_1} from the GAMA catalog \citep{Baldry12}, the IDL-based cosmological distance routine \texttt{lumdist.pro} \citep{Carroll92}, and the concordance cosmology \citep[H$_{0}$ = 70 km s$^{-1}$, $\Omega_{m}$ = 0.3, $\Omega_{\Lambda}$ = 0.7;][]{Hinshaw09}.  For SAMI cluster galaxies not included in GAMA, we calculate the distance using the host cluster redshift from \citet{Owers17}.

We convert the intrinsic \ha~luminosity maps into star formation rate maps using the conversion factor $7.9\times 10^{-42}$ M$_\odot$ yr$^{-1}$ (erg s$^{-1}$)$^{-1}$ from \citet{Kennicutt94}; to convert the assumed Salpeter initial mass function (IMF) to a \citet{Chabrier03} IMF, we divide the conversion by 1.53, following \citet{Gunawardhana13} and \citet{Davies16}.  We produce SFR maps in units of M$_\odot$ yr$^{-1}$ and SFR surface density maps in units of M$_\odot$ yr$^{-1}$ kpc$^{-2}$.  SFR surface density maps are not deprojected for inclination, although edge-on galaxies (i$>$80$^{\circ}$) are removed from the analysis.  By not deprojecting the surface densities, we absorb the systematic errors associated with a variety of inclinations: subsequent star formation surface density profiles of spiral galaxies can be affected by up to 0.5 dex.  We describe in each section how these systematics affect the interpretation of results.  

As an indirect optical tracer of the star formation rate, H$\alpha$ can suffer some systematic uncertainties due to dust and variations in the ionizing flux of massive stars. While we have corrected for extinction using the Balmer decrement, if the assumed extinction law is incorrect, or the optical H$\alpha$ line is so heavily obscured to be unobservable, H$\alpha$ will incorrectly measure the SFR. However, previous work from CALIFA has demonstrated that in resolved integral field studies, extinction corrected H$\alpha$ luminosities agree well with other SFR tracers \citep{Catalan-Torrecilla15}.  In Appendix~\ref{globalsfcorr}, we compare our SFR measurements with several multiwavelength SFR measurement techniques for the GAMA sample to confirm that \ha~is a reasonable tracer of SF in our sample.

The stochastic nature of star formation can severely bias some SFR tracers, particularly at low SFRs \citep{Cervino03, Cervino04, daSilva14}.
The bias for \ha~is minor for -4$<$log(SFR/M$_{\sun}$)$<$-3, but dramatically increases below this level.  About 25\% of the spaxels in this paper have \ha-predicted SFRs below 10$^{-4}$ M$_{\sun}$ yr$^{-1}$; however, all of these are eliminated by requiring an \ha~SNR of at least one and at least ten spaxels classified as star-forming within three effective radii.  We note that more moderate SFRs may still be subject to some bias; a full Bayesian analysis of that bias is beyond the scope of this work.

\subsubsection{Visual Morphological Classification}
\label{visualclassifications}

In our analysis, we also incorporate the visual classifications of morphology from \citet{Cortese14,Cortese16}.  These classifications follow the scheme used by the GAMA Survey \citep{Kelvin14}, first dividing galaxies into early- and late-types, and then splitting further based on the presence of a bulge.  We note that these classifications use Sloan Digital Sky Survey Data Release 9 \citep[SDSS DR9;][]{Ahn12} $gri$ three-colour images instead of the $giH$ images used by \citet{Kelvin14}.  
The SAMI classification scheme also differs slightly from that of \citet{Kelvin14} by additionally using colour to distinguish between Sa and S0 galaxies.  Because of these two changes, the final classifications from SAMI are on average shifted towards later galaxy types (see \citealt{Bassett17} for more details).  

\subsubsection{Stellar Mass Maps}
\label{stellarmassmaps}

For targets that are within the GAMA regions, we have access to high quality optical and NIR imaging from the VST KiDS \citep[$ugri$;][]{deJong15} and VIKING \citep[$zYJHKs$;][]{Edge13,Edge14} surveys.  These imaging data have been astrometrically matched to the same WCS as the SAMI cubes: i.e., the same 0.5\arcsec/pix scale, and the same nominal centre.  There are nevertheless minor astrometric mismatches between the multiband imaging and the SAMI cubes at the level of $\sim 0\farcs2$, due to, e.g., differential atmospheric refraction, pointing errors, etc. \citep{Green17}.  The imaging data have been smoothed to a consistent 1\arcsec~full-width at half-max point spread function (PSF); that is, the PSF is consistent across the multiwavelength imaging, but not necessarily between the imaging and the SAMI cubes.

We have derived stellar mass maps from these imaging data through stellar population synthesis (SPS) fits to the spectral energy distribution (SED) at each pixel location in these multiband image stacks. The SPS fitting process is the same as used for GAMA galaxies in GAMA DR3 \citep{Baldry17}, and closely follows \citet{Taylor11}. In brief, the fits combine the \citet{BruzualCharlot03} simple stellar population models with exponentially declining star formation histories, and single screen \citet{Calzetti00} dust. The fitting is done in a fully Bayesian way, with uniform priors on age, dust, and the e-folding time.  The nominal values for each fit parameter are the minimum mean square error (MSE) estimators, which are derived as PDF-weighted mean expectation values (see \citealt{Taylor11} for further discussion, and why it matters).

Where there is insufficient information in the SED to constrain the stellar population, there are a large number of models with very high mass-to-light ratios. This means that at low SNR, the $M/L$ can be systematically biased high.  The effect is driven primarily by the choice of uniform priors on dust: essentially, what happens is that the dust can take any (high) value, and the mass can similarly take any (high) value.  For this reason, we do not consider pixels where the combined SNR across the full SED is less than 10; simple tests suggest that this limits the bias in M/L to $\lesssim 0.1$ dex.  For combined SNR across the SED of 10, the formal uncertainty on the per-pixel stellar mass density is $\lesssim 0.25$ dex.  In practice, and with this SNR requirement, the pixel-to-pixel RMS in the inferred values of $M/L$ are $\lesssim 0.1$ dex.

\subsubsection{Environmental Density Measures}

In this paper, we make use of the environmental density measurements from the GAMA catalogue following the method outlined in \citet{Brough13}.  We choose to use the density defined by the fifth-nearest neighbour distance, $\Sigma_{5} = \frac{5}{\pi d_{5}^2}$, because simulations have shown that it is a good descriptor of the local dark matter density \citep{Muldrew12}.  The highly-complete ($>$95\%) GAMA-II survey is searched for neighbouring galaxies with absolute magnitudes M$_{r}<-18.5$ within $\pm$1000 km s$^{-1}$ to obtain the distance $d_{5}$ (in Mpc) to the fifth-nearest galaxy.  The resulting environmental density is scaled by the reciprocal of the survey completeness in that area, a correction of less than 5\%.  

$\Sigma_{5}$ can be biased when the edge of the survey footprint is closer than the fifth-nearest galaxy.  Such cases are flagged in the GAMA catalogue and are removed from this work.

\subsection{Subsamples Used in This Work}

The SAMI Galaxy Survey target selection was designed to probe a wide range of stellar masses and environments.  Here we discuss the properties of samples used in this work, which are drawn from the first 1296 SAMI galaxies (including duplicates), observed and reduced as of October 2015.

\begin{figure}
\centering
\includegraphics[scale=0.9,trim=0.8cm 0.6cm 0cm 0.8cm]{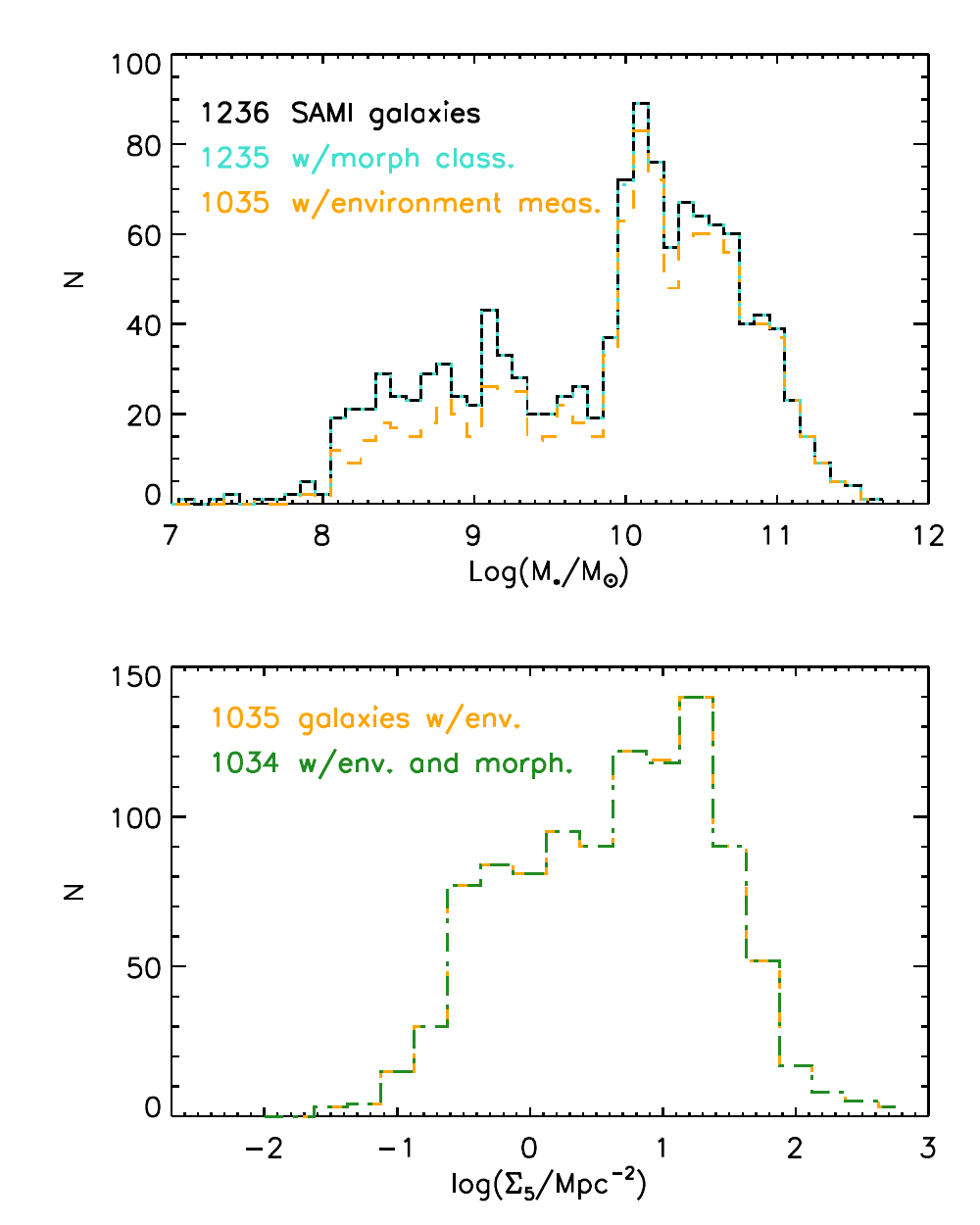}
\caption{\textit{Top:} Distribution of stellar masses of the SAMI sample (black solid line) used in the global star-forming main sequence work of Section~\ref{global}.  The orange dashed line shows the distribution of the subsample that has high-quality measurements of environmental density, $\Sigma_{5}$.  The turquoise dotted line shows the distribution of the sample for which morphological classifications are available.  \textit{Bottom:} Distribution of environmental densities of the sample of galaxies for which the measure is available (orange dashed line), and for the sample of galaxies that have both environmental densities and morphological classifications (green dot-dashed line).  The numbers in the top left correspond to the numbers of galaxies in each category.
} 
\label{hist_global}
\end{figure}

In Section~\ref{global}, we discuss the global star formation rates of our sample with respect to total stellar mass, environmental density, and morphology.  Figure~\ref{hist_global} shows the distributions in stellar mass and environmental density for the various subsamples we use in this work.  

\begin{figure}
\centering
\includegraphics[scale=0.9,trim=0.8cm 0.6cm 0cm 0.8cm]{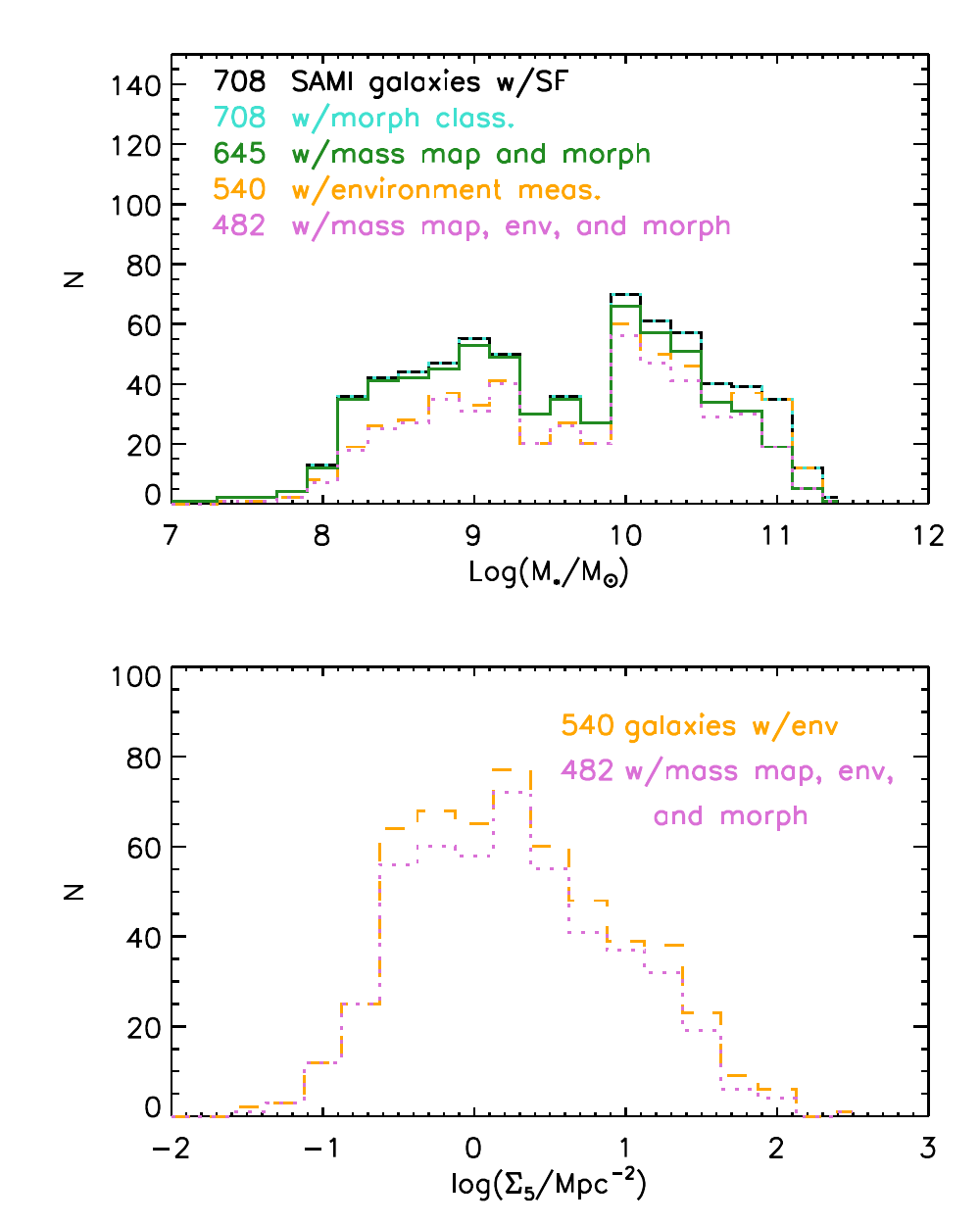}
\caption{
\textit{Top:} Distribution of stellar masses of SAMI sample (black solid line) used in the resolved star-forming main sequence work of Section~\ref{local}; these galaxies are the subset of those shown in Figure~\ref{hist_global} that have at least 10 spaxels of detectable star formation with SNR$>$1.  The turquoise dotted line shows the distribution of the subsample of resolved star formation that also has morphological classifications; the dark green solid line describes the subsample of resolved star formation with both morphological classifications and resolved stellar mass maps.
The orange dashed line shows the distribution of the subsample that has high-quality measurements of environmental density, $\Sigma_{5}$; the pink dotted line describes the subsample of galaxies that have all data products available.
\textit{Bottom:} Distribution of environmental densities of the samples of galaxies in the top panel for which the measure is available.  The numbers in the top left correspond to the numbers of galaxies in each category.
} 
\label{hist_local}
\end{figure}

In Section~\ref{local}, we discuss the resolved specific star formation rates and star formation rate surface densities of our sample with respect to morphology and environmental density.  Measuring specific star formation rates requires the spatially resolved stellar mass maps described in Section~\ref{stellarmassmaps}, which are only available for the GAMA sample; for completeness, however, we include galaxies without stellar mass maps in analyses of $\Sigma_{\text{SFR}}$ that do not explicitly require them.  In Figure~\ref{hist_local}, we show the properties of galaxies included in these various samples.
Although our requirements of high-quality environmental measures, morphological classifications, and stellar mass maps limit our sample, we see that the overall span of mass and environmental distributions is wide enough for an interesting investigation.  

\section{SAMI Galaxies on the Star-Forming Main Sequence}
\label{global}

\begin{figure}
\centering
\includegraphics[scale=0.73,trim=0.8cm 0.8cm 0cm 0.8cm]{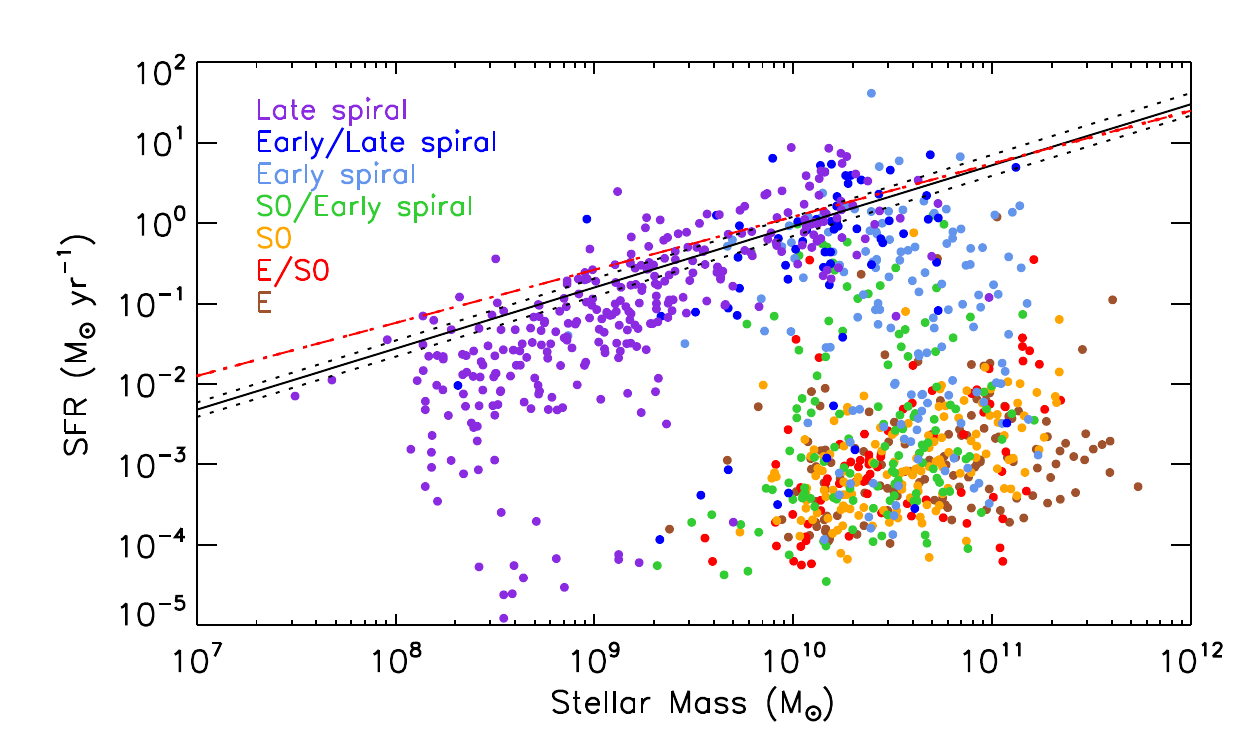}
\caption{Global star-forming main sequence of SAMI galaxies, split by morphological classifications from \citet{Cortese16}.  The black solid line shows the relation from \citet{RenziniPeng15}, the red dashed line shows the relation from \citet{Davies16}; the corresponding dotted lines show $\pm$1$\sigma$ for each relation.  SAMI galaxies do display a main sequence, evident mainly in spiral galaxies.  The slope of the sequence differs because SAMI star-formation rate maps are `clean': zeroed out where contamination from shocks or AGN might be contributing, or when emission lines are only marginally detected.  As a result, the SAMI total SFRs presented here should be considered lower limits.  See Appendix~\ref{globalsfcorr} for more details.} 
\label{globalSFMS_morph}
\end{figure}

We begin by comparing the integrated properties of SAMI galaxies to the global star-forming main sequence.  We sum the star formation rate maps for each galaxy to obtain the total star formation rate and use the total stellar masses from the GAMA catalog.  This global relation is shown in Figure ~\ref{globalSFMS_morph}, along with the fits to the main sequence from \citet{RenziniPeng15} and the u-band fit for all star-forming galaxies at 0$<$z$<$0.1 from \citet{Davies16}.  
SAMI galaxies visually classified as `late spirals' or `early/late spirals' follow a sequence, consistent with previous works, but with a steeper slope.  
In Appendix~\ref{globalsfcorr}, we make a detailed comparison of our global star formation rates to those from the GAMA survey \citep{Gunawardhana13, Davies16} to confirm that this steeper slope is a result of the star formation masks calculated in Section~\ref{SFmasks}, which produce a clean sample instead of a complete sample: that is, unless we are sure the spectrum is dominated by star formation, the spaxel is not counted.  Low SFR surface density regions may be below our detection limit for individual spaxels.  This mask will disproportionately affect galaxies at the lower end of our stellar mass range, as they are more likely to have low star formation rates and therefore only marginally-detected \ha~emission.  As a result, the global star formation rates calculated from SAMI presented here should be considered lower limits unless individually examined.  Note that, although the SAMI field-of-view covers most star formation for most galaxies, a true global star formation rate calculation should consider when aperture corrections are necessary \citep[see e.g.][for a discussion]{Green14, Richards16}.

\begin{figure}
\centering
\includegraphics[scale=0.73,trim=0.8cm 0.8cm 0cm 0.8cm]{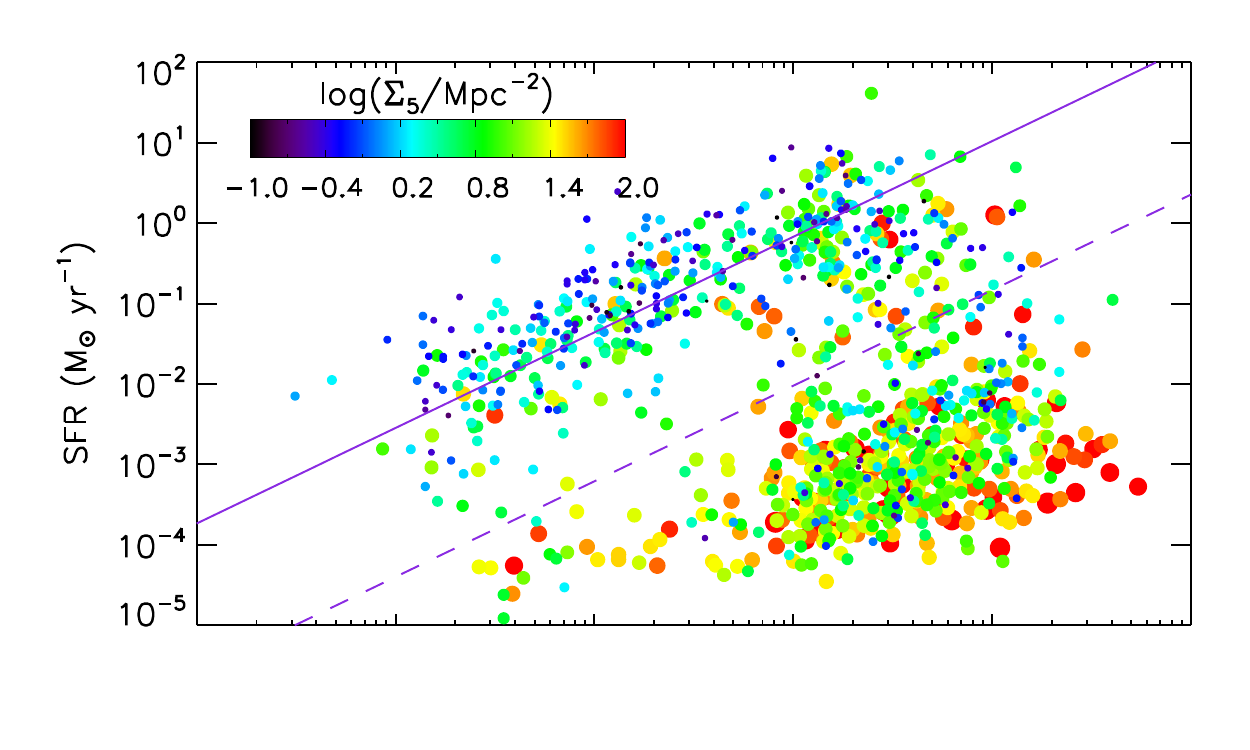}
\includegraphics[scale=0.73,trim=0.8cm 0.8cm 0cm 0.8cm]{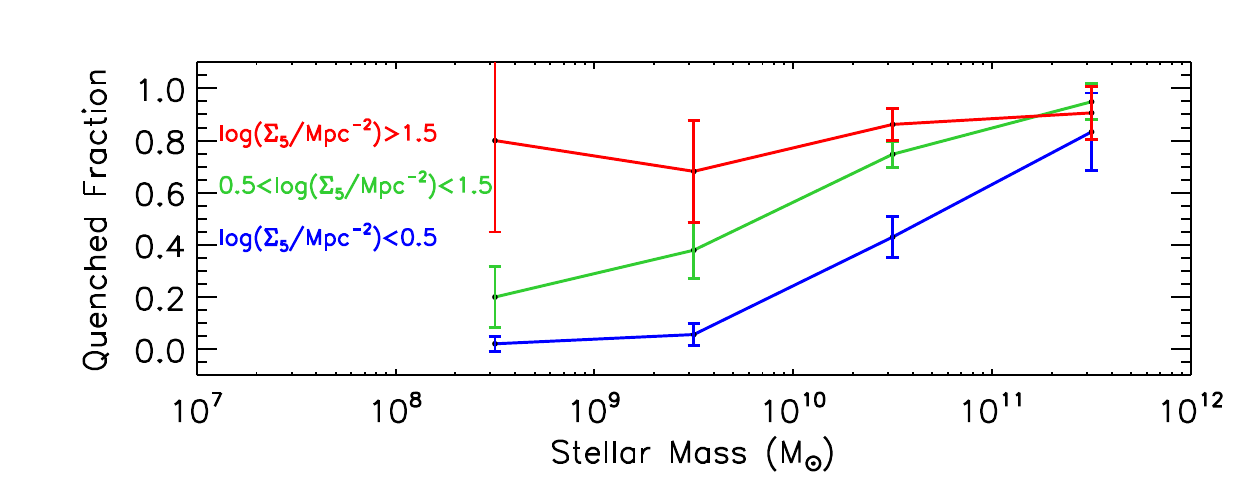}
\caption{\textbf{Top:} Global star-forming main sequence of SAMI galaxies, colour-coded by environmental density (isolated galaxies = small black points, galaxies in high density environments = large red points).  
We see a dearth of low-mass isolated galaxies below the main sequence: isolated high-mass galaxies can be quenched, but at masses $<10^{10}$ M$_{\sun}$, only galaxies in medium to dense environments are quenched.  Purple solid line shows a fit to the late-type spirals (purple points in Figure~\ref{globalSFMS_morph}); purple dashed line delineates galaxies that are 3$\sigma$ below the fit to late-type spirals.  
\textbf{Bottom:} Quenched fraction (fraction of galaxies below dashed purple line) as a function of stellar mass, divided into three environmental density bins.  Error bars show the 95\% confidence interval on the binomial fraction in each bin.  As above, galaxies in densest environments are more likely to be quenched at any mass; galaxies at higher masses are more likely to be quenched in any environment.
}  
\label{globalSFMS_env}
\end{figure}

One powerful aspect of the SAMI Galaxy Survey is the range of environments covered by the survey.  We therefore also plot the global star-forming main sequence colour-coded by environmental density (see Figure~\ref{globalSFMS_env}).  High-mass ($>10^{10}$ M$_{\sun}$) galaxies can lie on or below the main sequence at any environmental density, but low-mass ($<10^{10}$ M$_{\sun}$) galaxies are almost exclusively quenched in denser environments.  Note, however, that not all galaxies in dense environments are quenched.  These results are consistent with previous studies \citep[e.g.][]{Haines08,Geha12}.

\section{Spatially Resolving Galaxies on the Star-Forming Main Sequence}
\label{local}

The global star-forming main sequence compares the total star formation rate of a galaxy to its stellar mass, showing a nearly-constant sSFR between galaxies.  \citet{Cano-Diaz16} used CALIFA galaxies to show that individual regions of galaxies also follow a main sequence.

\begin{figure*}
\centering
\includegraphics[scale=0.8,trim=1cm 0.5cm 0cm 1cm]{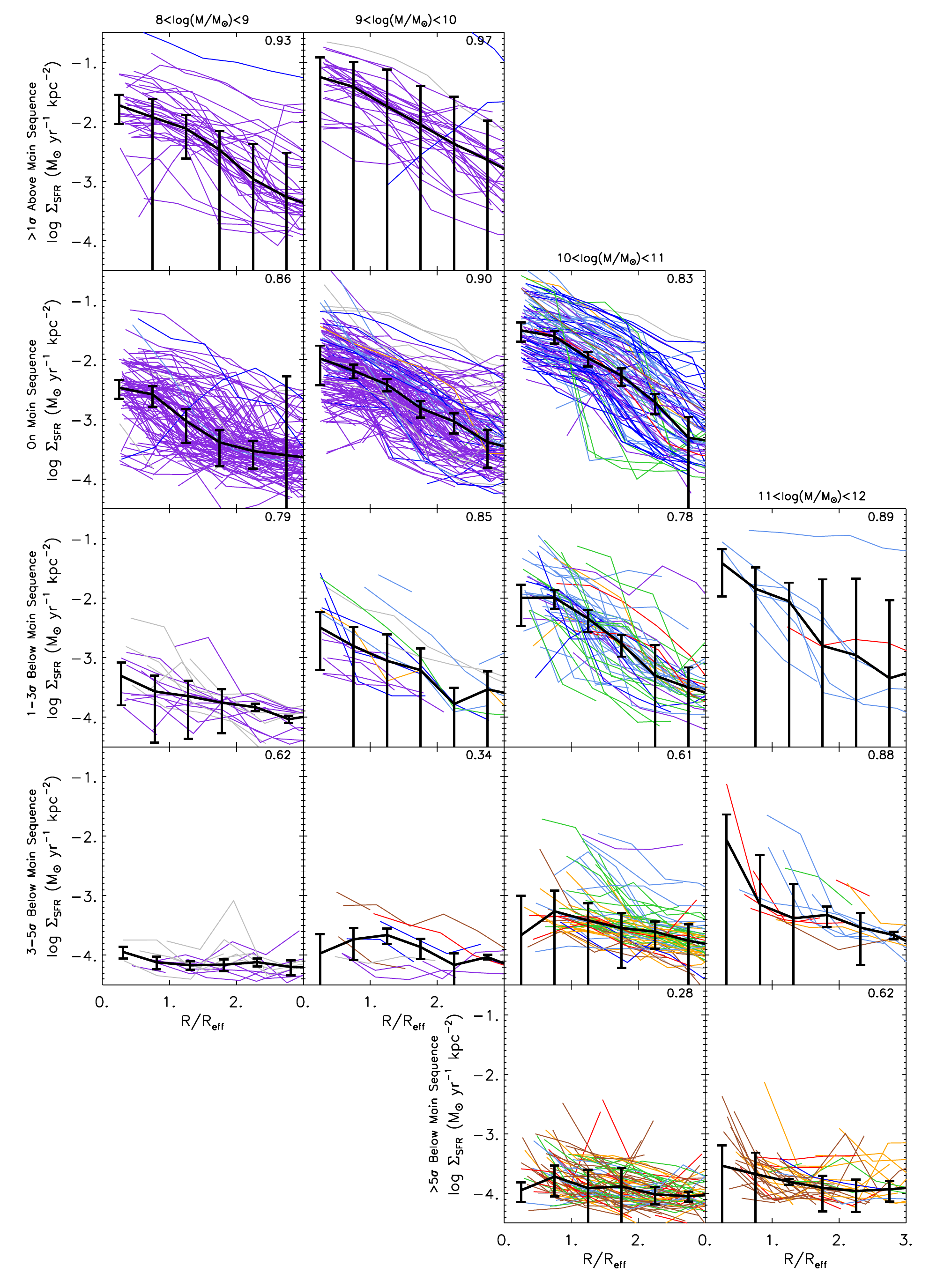}
\caption{
Radial profiles of star formation surface density for galaxies binned by stellar mass and location relative to the global star-forming main sequence.  The columns split galaxies into four bins of total stellar mass, denoted at the top of each column.  The rows split galaxies into five bins of integrated star formation rate defined relative to the main sequence fit in Figure~\ref{globalSFMS_env} ($>$1$\sigma$ above the main sequence, within 1$\sigma$ scatter of the main sequence, 1-3$\sigma$ below the main sequence, 3-5$\sigma$ below the main sequence, and $>$5$\sigma$ below the main sequence).  In each panel, the radial profiles (median of spaxels in each radial bin) of each galaxy are shown in the colour matching their morphological classification from Figure~\ref{globalSFMS_morph} or grey if visual classifications were unavailable, and the median radial profile across all galaxies as the thick black line; error bars are the standard error on the median, 
$1.253 \frac{\sigma_{\Sigma_{\text{SFR},n}}}{\sqrt{n}}$, 
where $\sigma_{\Sigma_{\text{SFR},n}}$ is the standard deviation of star formation rate surface densities for $n$ galaxies in a bin).  The completeness (fraction of galaxies in each bin of Figure~\ref{globalSFMS_env} for which we can measure $\Sigma_{\text{SFR}}$) is given in the top right of each panel.
} 
\label{SFRSDprofiles}
\end{figure*}

We use our integral field spectroscopy to examine the star formation rate surface density profiles $\Sigma_{\text{SFR}}$(r) of galaxies based on their location relative to the global star-forming main sequence.  To create these profiles, we calculate a galactocentric radius map using the inclination calculated from the ellipticity \citep{Bryant15} and fixing the centre of the galaxy to the peak of the stellar mass map from Section~\ref{stellarmassmaps}.  We then median-combine the $\Sigma_{\text{SFR}}$ of spaxels in bins of 0.5 R$_{\text{eff}}$ to create a radial profile for each galaxy.  Error bars for each bin are calculated using the standard error on the median, $1.253 \frac{\sigma_{\Sigma_{\text{SFR},n}}}{\sqrt{n}}$, where $\sigma_{\Sigma_{\text{SFR},n}}$ is the standard deviation of star formation rate surface densities for $n$ galaxies in a bin.  When combining multiple galaxies into a median profile, we first bin each individual galaxy and then median-combine the binned galaxy profiles, to ensure that closer galaxies (with more spaxels per bin) are not more heavily weighted.  Each panel of Figure~\ref{SFRSDprofiles} shows the completeness of the bin in the top right corner.  Of the 217 galaxies in our sample excluded from this figure, 23 (11\%) had strong \ha~detections but were identified as largely contaminated by shocks/AGN (see Section~\ref{SFmasks}) and 194 (89\%) may have had star formation below our detection limits.
We note that bins with low completeness likely show overestimates of the median profiles, because the missing galaxies likely have lower $\Sigma_{\text{SFR}}$ than the detected galaxies.  However, because strong detections rely on the surface brightness of \ha~relative to the stellar continuum, each galaxy has a different detection limit.  The $\Sigma_{\text{SFR}}$(r) profiles of spiral galaxies are not corrected for inclination, and can therefore be inflated relative to their face-on counterparts by up to 0.5 dex: the inclination effect likely drives some of the variation in the profiles of spiral galaxies.  However, because our galaxies are well distributed in inclination, the median profiles are likely only affected by $<$0.2 dex, a small effect compared to our uncertainties.

Figure~\ref{SFRSDprofiles} shows $\Sigma_{\text{SFR}}$ profiles for our galaxies in four bins of stellar mass and five bins of vertical distance from the main sequence fit of Figure~\ref{globalSFMS_env}.  As expected, the overall $\Sigma_{\text{SFR}}$ levels increase with higher stellar mass and decrease with distance below the main sequence.  Galaxies on and around (within 3$\sigma$ of) the main sequence on average have centrally-concentrated star formation: higher star formation rate surface densities in the nuclei.  Individual galaxies in these bins do still show considerable scatter, both in overall $\Sigma_{\text{SFR}}$ levels and in the shape of the radial profiles.  This scatter is present at all radii: starbursting galaxies are not merely normal galaxies with a bright burst of nuclear star formation.

When controlling for mass, galaxies lying below the main sequence show progressively flatter $\Sigma_{\text{SFR}}$ profiles, suggesting that the bulk of the quenching happens from the inside out.  We note, however, that even 3-5$\sigma$ below the main sequence (i.e. fourth row), some galaxies show centrally-concentrated $\Sigma_{\text{SFR}}$ profiles while some galaxies show completely flat profiles.  Our lowest $\Sigma_{\text{SFR}}$ bin (bottom row) shows mainly galaxies with considerably flatter profiles.  

\citet{Nelson16} used a stacking analysis of 3D-HST observations to show that galaxies within 2$\sigma$ of the main sequence have centrally-concentrated SFR profiles and generally-flat sSFR profiles.  
However, from our analysis it is clear that the strongest cases of quenching occur in galaxies $>$3 sigma below the main sequence. Therefore, the smaller range probed by \citet{Nelson16} instead reflects only minor oscillations in star formation activity \citep[such as those predicted by][]{Tacchella16a,Tacchella16b}.

In the sections below, we probe the star-forming main sequence and the sSFR profiles within individual galaxies further: as a galaxy quenches, does the sSFR gradually decrease across the entire galaxy, or do some regions maintain the typical sSFR while others shut off completely?  We examine the star-forming main sequence using the star formation rate surface densities ($\Sigma_{\text{SFR}}$ in M$_{\sun}$ yr$^{-1}$ kpc$^{-2}$) and stellar mass densities ($\Sigma_{M_*}$ in M$_{\sun}$ kpc$^{-2}$), and calculate the sSFR in individual spaxels (SFR maps in M$_{\sun}$ yr$^{-1}$ divided by stellar mass maps in M$_{\sun}$).
We examine the impact of morphology, environmental density, and stellar mass on these relations.

\subsection{Morphological Quenching}
\label{morph}
\subsubsection{Morphological Effects on the Resolved Star-forming Main Sequence}
\begin{figure*}
\centering
\includegraphics[scale=0.95,trim=1cm 1cm 0cm 0.5cm]{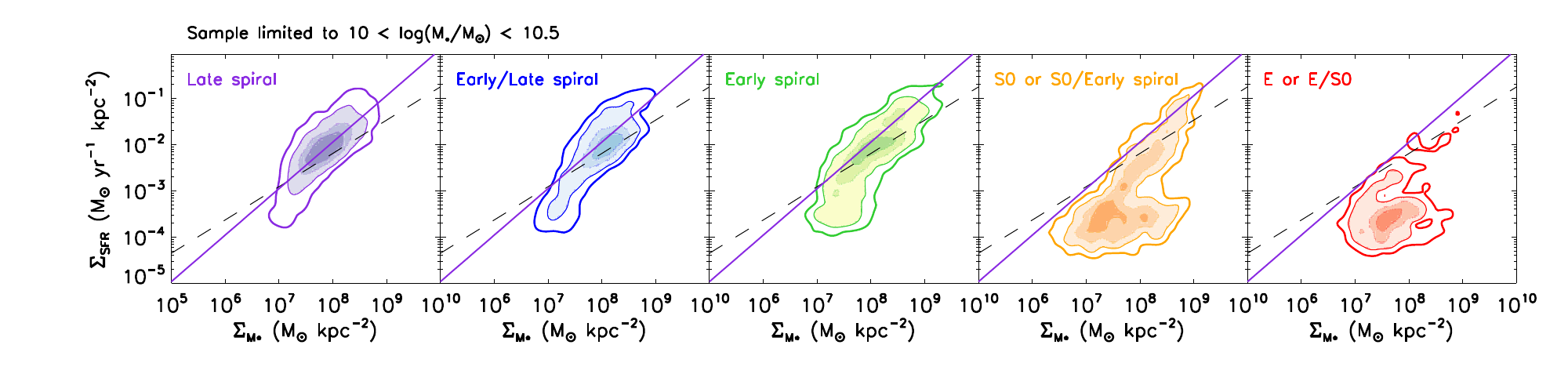}
\caption{
The resolved star-forming main sequence: star formation rate surface density ($\Sigma_{\text{SFR}}$ in M$_{\sun}$ yr$^{-1}$ kpc$^{-2}$) as a function of stellar mass surface density ($\Sigma_{M_*}$ in M$_{\sun}$ kpc$^{-2}$) for SAMI galaxies with 10$<$log(M$_{*}$/M$_{\sun}$)$<$10.5 divided into different morphological categories according to \citet{Cortese16}.  Each panel shows contours of the 2D histogram of spaxels on the sequence for galaxies of the noted morphological type.  Contour levels are at 10\%, 25\%, 50\%, and 75\% of the maximum density levels.  Each panel shows the fit to the star-forming main sequence from \citet[][black dashed line]{Cano-Diaz16} and the fit to the late spiral galaxies shown here in the left panel (purple solid line).  Only our two earlier-type morphological bins of SAMI galaxies show a ridgeline resembling the relation from \citet{Cano-Diaz16}, which is to be expected from their sample demographics.  All morphological categories contain some spatial regions that match the late-type spiral relation, but earlier-type galaxies show an increasing quantity of spaxels lying below the relation.
} 
\label{localSFMS_morph}
\end{figure*}

We use our SAMI integral field spectroscopy to examine the spatially-resolved star-forming main sequence of galaxies across a range of morphological types (Figure~\ref{localSFMS_morph}).  Like \citet{Cano-Diaz16}, we see a strong relation between star formation rate surface density and stellar mass surface density, particularly in later-type spiral galaxies.  However, our relation for late-type spiral galaxies is steeper (slope of $1.00\pm0.01$ as compared to their $0.72\pm0.04$).  This discrepancy is likely due to the range of morphological types included in the CALIFA fit; indeed, the upper ridgelines of early-type galaxies (orange and red, fourth and fifth panels) resemble their fit.  Neither surface density is corrected for projection effects, which we expect to spread our distributions out along the main sequence, rather than above/below it; projection effects do not affect our interpretation of this and similar plots.

When splitting by morphological types, we find that our earlier-type galaxies contain two kinds of spatial regions.  The upper regions form a main sequence similar to that of late-type spirals.  This sequence is offset towards lower $\Sigma_{\text{SFR}}$ and/or higher $\Sigma_{\text{M}*}$, although the spread is such that some regions are forming stars at the rate predicted by the late-type spiral main sequence.  The other regions drop to lower $\Sigma_{\text{SFR}}$ values.  
We note that these spaxels are not simply \ha~emission caused by the diffuse ionized medium or old stars \citep[see e.g.][]{JBH91, Binette94}: when limiting our analysis to spectra with \ha~equivalent widths $<$-3\AA~(where negative is emission), the lower populations remain in the two early-type bins.  We investigate these two populations further in Section~\ref{splitpops}.

The distributions of star formation rate surface density relative to stellar mass surface density are statistically distinct for the different morphological classes.  Two-sided Kolmogorov-Smirnov and Anderson-Darling tests find p-values of $<$0.01 when comparing the sSFR distributions of late-type spirals within a 0.5 dex stellar mass surface density bin to the corresponding distribution of all other morphological classes.

\subsubsection{The Radial Component of Morphological Effects}
\citet{GonzalezDelgado16} have shown that quenching occurs from the inside-out in the CALIFA sample: early-type galaxies have lower sSFRs in their nuclear regions than in their outskirts.  \citet{Belfiore17a} drew a similar conclusion for MaNGA galaxies, finding that galaxies with intermediate-age stellar populations in the nuclei lie just below the star-forming main sequence, while galaxies with widespread intermediate-age stellar populations lie significantly below the main sequence, akin to quiescent galaxies.  These centrally-quenching galaxies lie in the green valley and are associated with the build-up of a stellar bulge \citep{Belfiore17b}.

\begin{figure*}
\centering
\includegraphics[scale=0.95,trim=1cm 0.8cm 0cm 0.8cm]{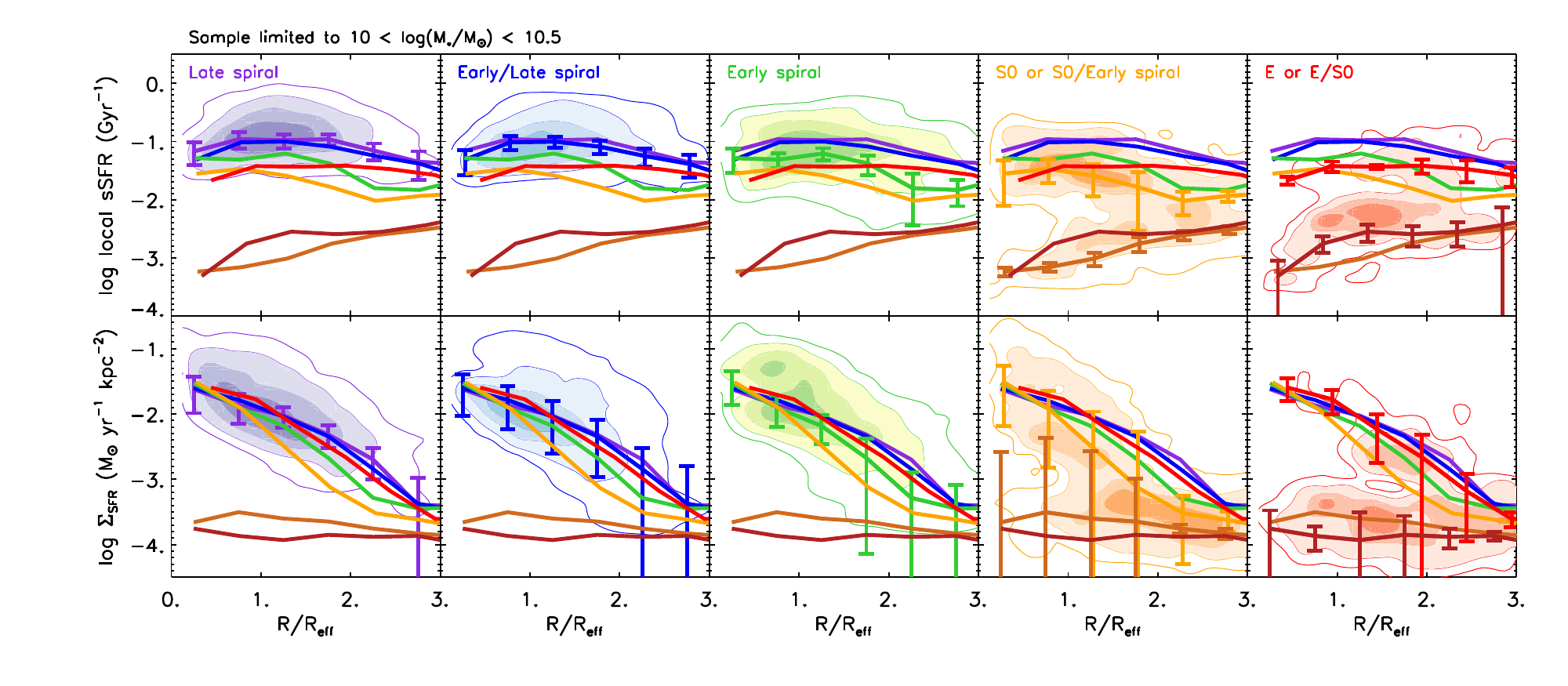}
\caption{
\textbf{Top row:} Radial profiles of specific star formation rate per spaxel in units of Gyr$^{-1}$ for SAMI galaxies with 10$<$log(M$_{*}$/M$_{\sun}$)$<$10.5 divided into different morphological categories according to \citet{Cortese16}.  Each panel left-to-right shows radial profiles of all morphological classes and contours showing the distribution of a single class.  Each radial bin shows the median-level sSFR across galaxies in that sample; error bars are the standard error on the median, 
$1.253 \frac{\sigma_{\Sigma_{\text{SFR},n}}}{\sqrt{n}}$, 
where $\sigma_{\Sigma_{\text{SFR},n}}$ is the standard deviation of star formation rate surface densities for $n$ galaxies in a bin).
`S0 or S0/Early spiral' (orange/brown) and `E or E/S0' (red/dark red) galaxies show split populations (Section~\ref{splitpops}) and show two radial profiles each.  
\textbf{Bottom row:} Radial profiles of star formation rate surface density in units of M$_{\sun}$ yr$^{-1}$ kpc$^{-2}$ for the same samples as top row.  
Moving to the right to earlier-type galaxies and including only the upper populations of early-type galaxies, we see flat sSFR profiles with slightly lower sSFR values at earlier types and centrally-concentrated $\Sigma_{\text{SFR}}$ profiles that do not vary between morphological types.  
The decreased sSFR in the lower population of early-type galaxies is not solely due to a build-up of mass because the same galaxies show a decreased $\Sigma_{\text{SFR}}$ profile as well.
}
\label{localSSFR_morph}
\end{figure*}

We present in Figure~\ref{localSSFR_morph} an analogous approach using SAMI data, limiting our sample to galaxies with 10$<$log(M$_{*}$/M$_{\sun}$)$<$10.5 to remove mass effects.  
In each panel, the contours indicate the distribution of spaxels for a given morphological classification, with median profiles of all classifications overlaid for comparison.  As in Figure~\ref{SFRSDprofiles}, median profiles are first calculated for each galaxy and then combined to obtain the profile of the median galaxy.
SAMI galaxies also show a decreased nuclear and overall sSFR in the earliest-type galaxies.  Our morphological classification scheme does not explicitly separate all Hubble types, but we do see a decrease in overall sSFR as we move from late-type spirals to earlier spirals.  As in the previous subsection, we expect minimal effect on sSFR (top row) due to projection because both the SFR and stellar mass surface densities are uncorrected.  However, the $\Sigma_{\text{SFR}}$ profiles may be inflated by up to 0.5 dex between categories if the inclination distributions also vary.
Although \citet{GonzalezDelgado16} divide CALIFA galaxies into a different set of morphological categories, our resulting profiles show broadly consistent results.  The underlying distribution of sSFRs (filled contours) show that the dominant effect bringing down early-type galaxies is the presence of a quenched population: rather than the entire population moving to much lower sSFRs, early-type galaxies split into quenched and non-quenched subpopulations.  The second row of Figure~\ref{localSSFR_morph} shows that this lower sSFR is explicitly due to a population with lower $\Sigma_{\text{SFR}}$, rather than simply additional stellar mass washing out the sSFR profile.  We explore these two populations in the following section.

The distributions of the radial profiles of early-type (orange and red) galaxies are statistically distinct from our sample of late-type spirals.  When splitting the points into bins of 0.5 R$_{\text{eff}}$, two-sided 
Kolmogorov-Smirnov and Anderson-Darling tests find p-values of $<$0.01 when comparing the sSFR and $\Sigma_{\text{SFR}}$ distributions to the corresponding distribution in late-type spirals.
The differences in distributions may be even higher than shown by the profiles detected here because we ignore galaxies and spaxels that have star formation below our detection limit.  In this mass bin, we detect $>$85\% of galaxies of each morphology, but the detection coverage across individual galaxies is sometimes low, suggesting that individual profiles could be overestimated in some cases.  However, the lower population galaxies on average have fewer detected spaxels than the upper galaxies, so any overestimation of SF happens in the quenched population and we know the split is real.

\subsubsection{The Split Populations of Early-Type Galaxies}
\label{splitpops}

\begin{figure}
\includegraphics[scale=0.8,trim=1.5cm 1.2cm 0cm 0.8cm]{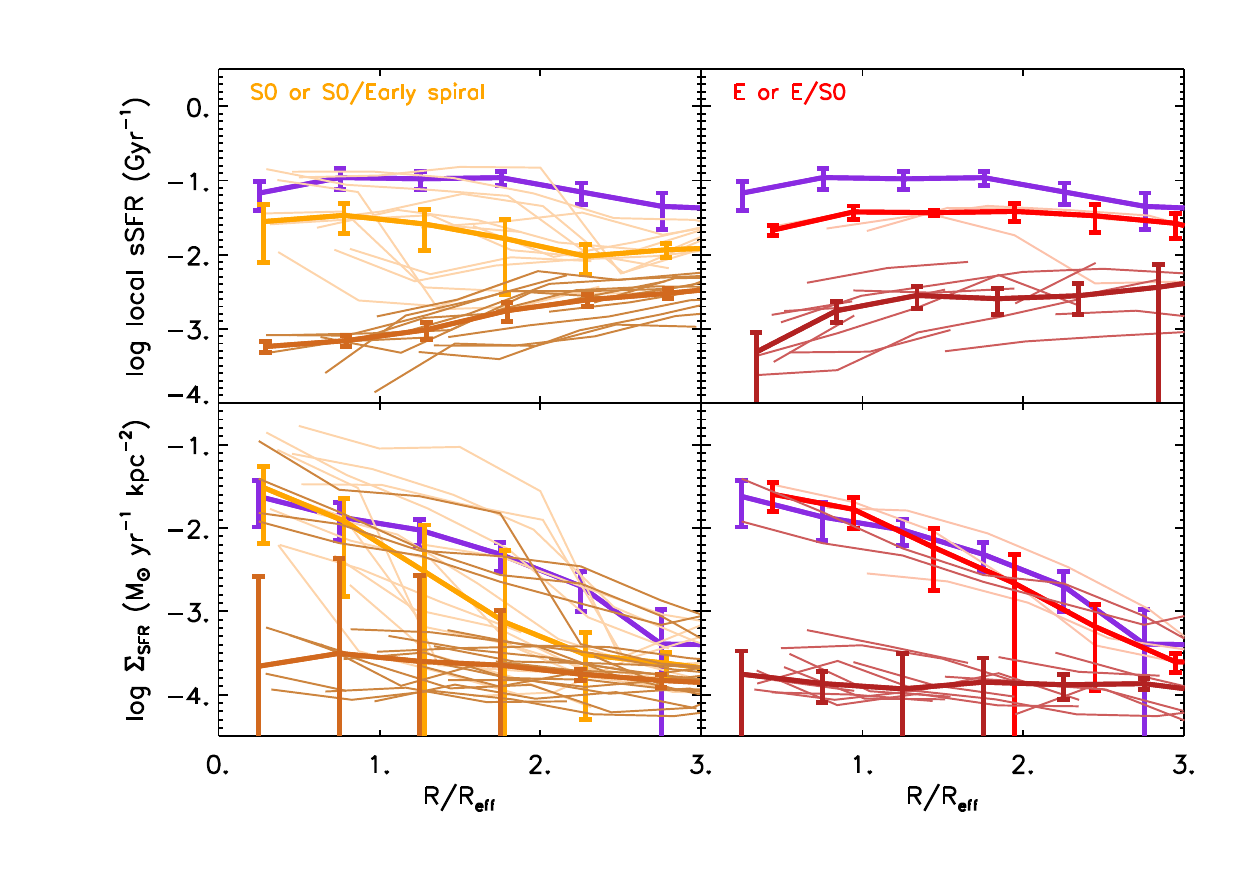}
\caption{
\textbf{Top row:} Radial profiles of specific star formation rate per spaxel in units of Gyr$^{-1}$ for SAMI galaxies with 10$<$log(M$_{*}$/M$_{\sun}$)$<$10.5 as in Figure~\ref{localSSFR_morph}, showing only early-type morphological categories according to \citet{Cortese16}.  Each panel shows the radial profiles of individual galaxies (thin lines), the overall median profile of late-type spirals (thick purple lines) for comparison, and the median profiles of the main sequence (orange and red thick lines) and below main sequence (brown and dark red thick lines) populations.  
\textbf{Bottom row:} Radial profiles of star formation rate surface density in units of M$_{\sun}$ yr$^{-1}$ kpc$^{-2}$ for the same samples as top row. 
The main sequence populations of both morphological categories mimic the star formation profiles of late-type galaxies: flat sSFR profiles (at a decreased level relative to late-type spirals) and centrally-concentrated $\Sigma_{\text{SFR}}$ profiles.  Most galaxies below the main sequence are similar between the `S0 or S0/Early spiral' (brown) and `E or E/S0' (dark red) galaxies and quite distinct from late-type galaxies: centrally-depressed sSFR profiles and flatter $\Sigma_{\text{SFR}}$ profiles.
} 
\label{earlytype_splitpops}
\end{figure}

Our two early-type morphological bins `S0 or S0/Early spiral' (orange) and `E or E/S0' (red) show a split in sSFR and $\Sigma_{\text{SFR}}$ (Figures~\ref{localSFMS_morph}~and~\ref{localSSFR_morph}).  These subpopulations represent two sets of galaxies with different SF behaviours as opposed to two sets of spaxels. In fact, the upper population exclusively comes from galaxies within 3$\sigma$ of the global SFMS (i.e. above the purple dashed line in the top panel of Figure~\ref{globalSFMS_env}).
In Figure~\ref{earlytype_splitpops}, we show radial profiles for the individual galaxies in these two morphological bins alongside the overall median profiles of each category's two subpopulations.  In both cases, the main sequence galaxies follow similar SF behavior to late-type galaxies: a flat (or centrally-concentrated) sSFR profile at a level slightly decreased relative to later-type galaxies and a centrally concentrated $\Sigma_{\text{SFR}}$ profile.  Galaxies $>$3$\sigma$ below the main sequence instead show a drastically different set of profiles: flat or centrally-depressed sSFR profiles and flat $\Sigma_{\text{SFR}}$ profiles.

\begin{figure}
\includegraphics[scale=0.85,trim=1cm 1.2cm 0cm 0.8cm]{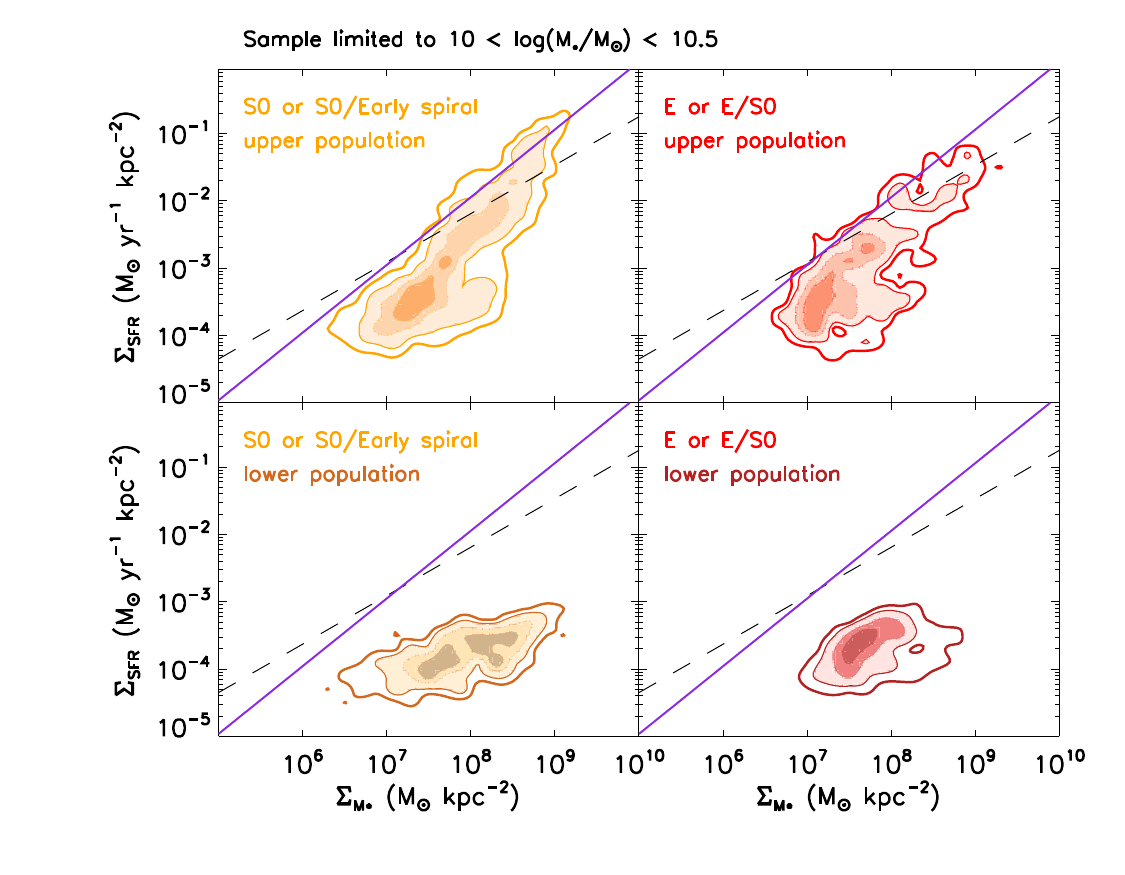}
\caption{
The resolved SFMS as in Figure~\ref{localSFMS_morph} for the split populations of early-type galaxies. Galaxies on the global SFMS follow a similar resolved main sequence to late-type galaxies, although offset down or to the right, consistent with the idea that bulges add additional stellar mass without affecting SF in these galaxies.  Galaxies below the global SFMS follow a lower, shallower trend in which star formation is significantly decreased.  } 
\label{splitpops_localSFMS}
\end{figure}

We additionally compare the spaxel-by-spaxel local SFMS for galaxies on and below the main sequence in Figure~\ref{splitpops_localSFMS}.  Splitting the populations in this frame confirms that main sequence population galaxies fill positions similar to late-type galaxies, although offset such that a particular $\Sigma_{\text{SFR}}$ value occurs at a higher $\Sigma_{\text{M}*}$ in S0s than it would in late-type spirals.  
Galaxies below the main sequence follow a trend with a much shallower slope, with increasingly different $\Sigma_{\text{SFR}}$ from main sequence galaxies at higher $\Sigma_{\text{M}*}$.

\begin{figure}
\includegraphics[scale=0.65,trim=0cm 0cm 0cm 0cm]{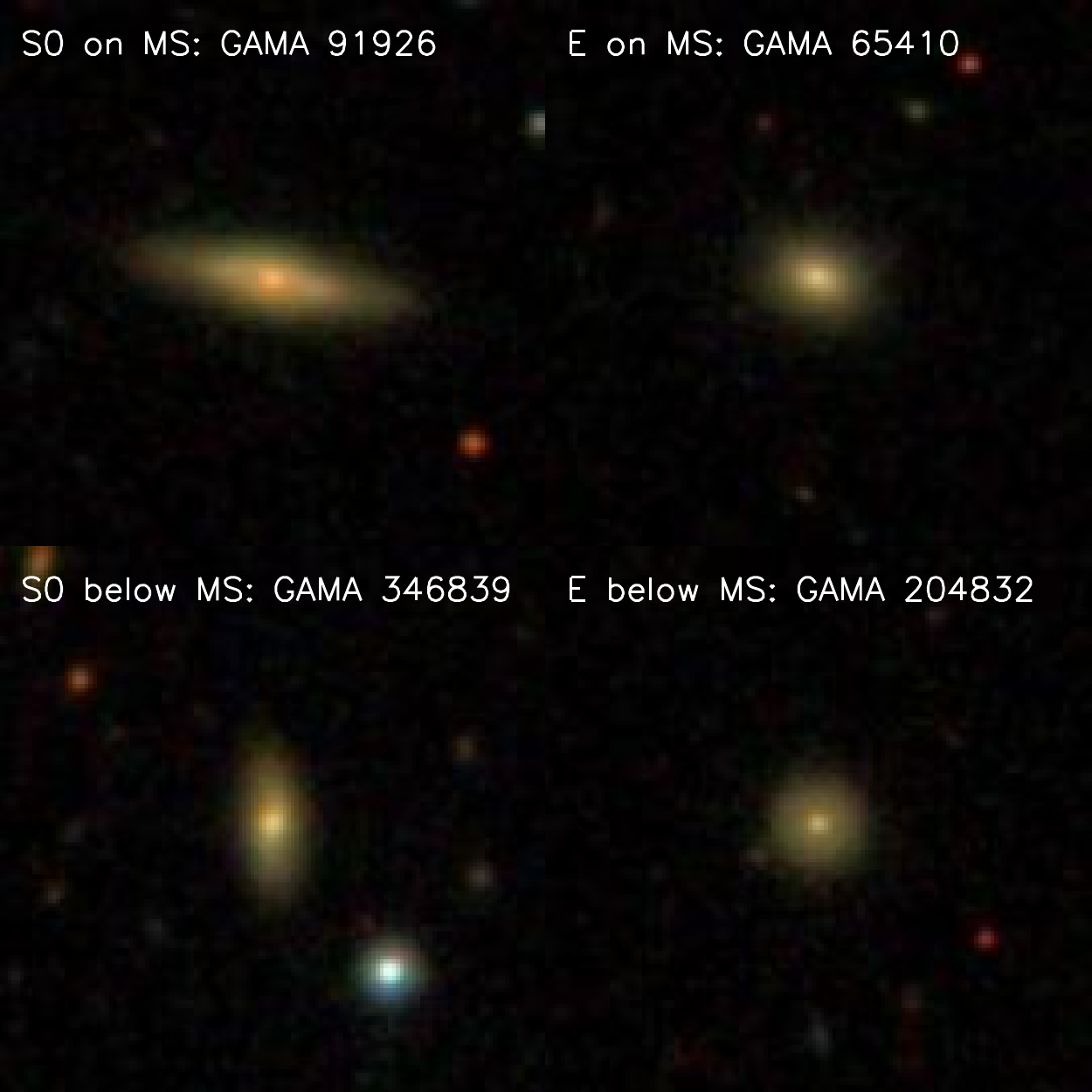}
\caption{
Three-colour SDSS images used for morphological classifications of four example galaxies from Figure~\ref{earlytype_splitpops}, one from each category.  Visually, galaxies in the upper and lower populations are indistiguishable.} 
\label{splitpops_3color}
\end{figure}

Early-type galaxies on the global main sequence are not simply late-type galaxies mislabelled by our visual morphological classification.  Figure~\ref{splitpops_3color} shows 3-colour images of a typical example from each category and confirms that there is little morphological difference between early-type galaxies on and below the main sequence.
The differences in SAMI morphological classification compared to those of the GAMA Survey would also not produce main sequence early-type galaxies: our classifications may move star-forming early spirals to late spirals, not the other way around.
The images in Figure~\ref{splitpops_3color} also suggest against the scenario that early-type galaxies by default fall below the main sequence and that bars or recent galaxy interactions have moved some to the main sequence.  The two populations differ in global SFR by $>$0.1 M$_{\sun}$ yr$^{-1}$; simulations suggest that a satellite with gas mass $10^{8}$ M$_{\sun}$ can stimulate new SF up to 0.005 M$_{\sun}$ yr$^{-1}$ \citep{Mapelli15}.  A satellite large enough to inject enough SFR to move a galaxy to the main sequence would therefore be visible in the classification images.  

Visual inspection of these SDSS images reveals no increased incidence of bars in either subset of early-type galaxies.  Only one of these galaxies has archival imaging from the Hubble Space Telescope.  This galaxy, GAMA 517302, resides in the main sequence S0 population; the high resolution imaging may show a sign of a weak bar undetectable in the SDSS image.  However, \citet{Ellison11} found that bars increase global SFRs by $\sim$60\% on average, whereas our main sequence population galaxies have 5-100 times higher SFRs than the lower population in the same mass bin.  We therefore don't think that bars are entirely responsible for the split between these two populations.  Further, the galaxies in our sample below the main sequence have $\sim$100 times higher SFRs than galaxies below the GAMA detection limit, so undetected bars or minor mergers would be insufficient to nudge completely quiescent galaxies up into this population.


\begin{figure}
\includegraphics[scale=0.9,trim=1cm 0cm 0cm 0cm]{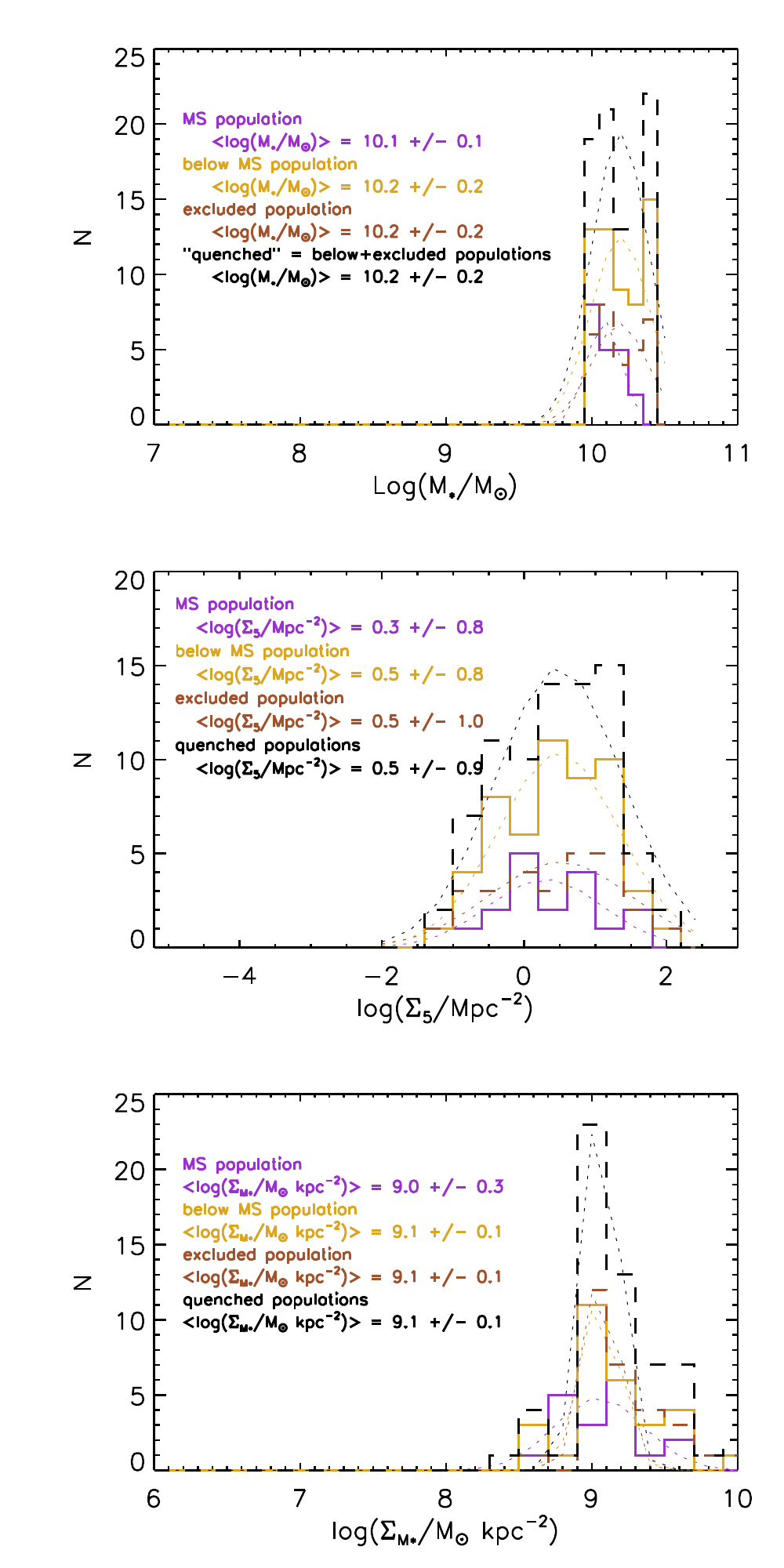}
\caption{
Histograms showing the distributions in total stellar mass (top), environment (middle), and central stellar mass surface density (bottom) of the main sequence (purple) and below main sequence (gold) populations.  For clarity, we also show the histograms of the galaxies excluded from our sample (brown dashed) and a putative ``quenched'' population -- the population we detect $>$3$\sigma$ below the main sequence plus those excluded, which may have star formation below our detection limit (black dashed).  In each case, a histogram is fit to the distribution and the average and Gaussian $\sigma$ are given in the appropriate panel.  With our sample size, no significant difference in any of these parameters is seen between any of these populations. }
\label{splitpops_histograms}
\end{figure}

Because morphological differences do not determine whether early-type galaxies fall on or below the global SFMS, we consider other possible drivers.  In Figure~\ref{splitpops_histograms}, we show the distributions of stellar mass, environment, and central stellar mass surface density for the main sequence and below main sequence populations.  We also include a ``quenched'' population that includes galaxies we detect below the main sequence plus all galaxies excluded from our sample (i.e. perhaps because their levels of SF are below our detection limit).  We see no statistically significant difference in any of these properties between the main sequence or below (or total quenched) populations, although galaxies on the main sequence have slightly lower stellar masses, live in slightly less dense environments, and have slightly lower central stellar mass surface densities.  A larger sample would confirm whether these distributions are actually distinct, but our sample definitively shows that there is no hard cutoff in stellar mass, environment, or central stellar mass surface densities where a galaxy drops off the main sequence.  We note that our two populations (and the excluded galaxies we include for comparison) are limited in stellar mass already to 10$<$log(M$_*$/M$_{\sun}$)$<$10.5.  It is therefore not surprising that the overall stellar mass distributions are not significantly different.  If we compare the distribution of main sequence (upper) and below main sequence (lower/quenched) early type galaxies without a mass limit, the differences in each parameter grow, but are still not statistically significant.

The stark differences between early-type galaxies on and below the global SFMS suggest that the presence of a bulge alone is not sufficient to quench galaxies; some of these galaxies with bulges continue their star formation exactly as late-type spirals would.  Indeed, a substantial fraction of early-type galaxies contain ongoing SF \citep[][and references therein]{Sarzi08}.  The main sequence populations show SF continuing in the disc without effect from the growing bulge; the bulge merely serves to decrease the sSFR slightly by adding mass without new SF.  Galaxies below the main sequence, on the other hand, exhibit significantly diminished SF in the nuclear regions that cannot be solely linked to stellar mass, environmental density, or central stellar mass surface density.  In Section~\ref{discussion}, we will explore possible causes for the split.

\subsection{Environmental Quenching}
\label{environment}
\subsubsection{Environmental Effects on the Resolved Star-forming Main Sequence}

\begin{figure*}
\centering
\includegraphics[scale=0.9,trim=1cm 0.5cm 0cm 0.5cm]{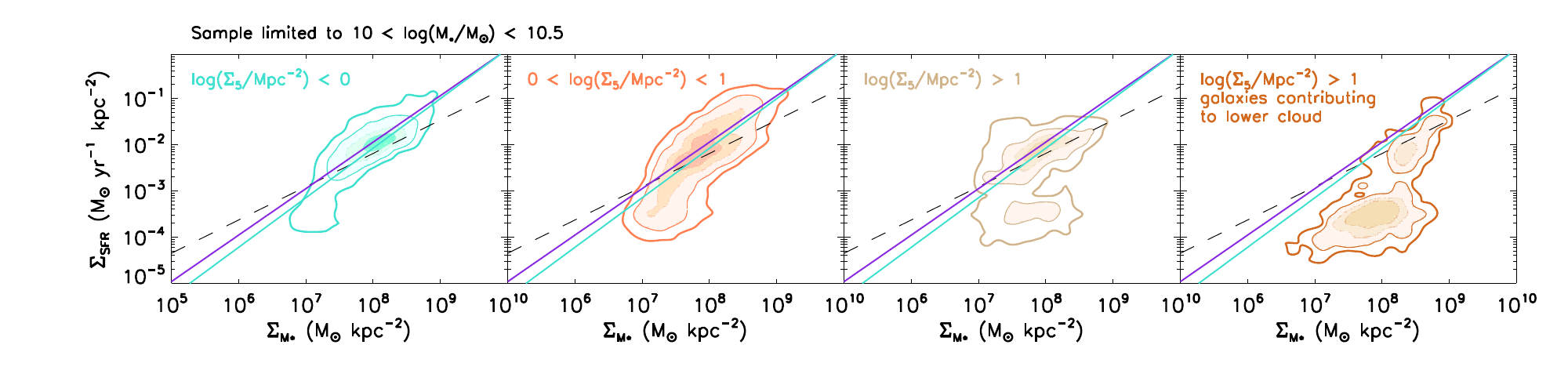}
\caption{
The resolved star-forming main sequence: star formation rate surface density ($\Sigma_{\text{SFR}}$ in M$_{\sun}$ yr$^{-1}$ kpc$^{-2}$) as a function of stellar mass surface density ($\Sigma_{M_*}$ in M$_{\sun}$ kpc$^{-2}$) for SAMI galaxies with 10$<$log(M$_{*}$/M$_{\sun}$)$<$10.5 divided into three environmental density bins according to \citet{Brough13}.  The fourth (rightmost) panel shows a subset of galaxies from the high density bin: those that have $\Sigma_{\text{SFR}}<10^{-3}$ M$_{\sun}$ yr$^{-1}$ kpc$^{-2}$ in $>$25\% of their spaxels.
Each panel shows contours of the 2D histogram of spaxels on the sequence for galaxies in the noted environmental density range.  Contour levels are at 10\%, 25\%, 50\%, and 75\% of the maximum density levels.  Each panel shows the fit to the star-forming main sequence from \citet[][black dashed line]{Cano-Diaz16} and the fit to the late spiral galaxies shown in the left panel of Figure~\ref{localSFMS_morph} (purple solid line).  We also show the fit to all low-density galaxies from the left panel (cyan solid line), which is very similar to the late-type spiral fit.  
In dense environments, the galaxies making up the bulk of the lower cloud also have regions that fall on or near the SFMS-level.
} 
\label{localSFMS_env}
\end{figure*}

As discussed in Section~\ref{global}, the environment in which a galaxy lives likely plays a role in quenching.  We show in Figure~\ref{localSFMS_env} the spatially-resolved star-forming main sequence of galaxies split into three bins of environmental density according to \citet{Brough13}.  We see minimal change in the upper envelope of the sequence with environmental density.  However, the tail of spaxels falling below the main sequence becomes more prominent at medium- and high-densities (log($\Sigma_{5}$/Mpc$^{-2})>0$, $>1$).  This tail is sufficient to make the $\Sigma_{\text{SFR}}$ distributions of galaxies in medium- and high-density regions statistically distinct from those in low-density regions even when controlling for stellar mass surface density: the two-sided Anderson-Darling and Komolgorov-Smirnov p-values comparing the distributions in stellar mass surface density bins of 0.5 dex are all $<$0.01.  To explore this tail of low SFR surface density regions in the high density bin, we examine the galaxies that contribute most heavily.  Unlike in Section~\ref{splitpops}, where the early-type galaxies split cleanly into two populations near and below the local SFMS, here the galaxies falling below also contain regions on or near the local SFMS.  These normal star-forming regions tend to be at higher stellar mass surface densities than those regions experiencing some quenching.

\subsubsection{The Radial Component of Environmental Effects}

\begin{figure*}
\centering
\includegraphics[scale=0.9,trim=0.8cm 0.3cm 0cm 0.8cm]{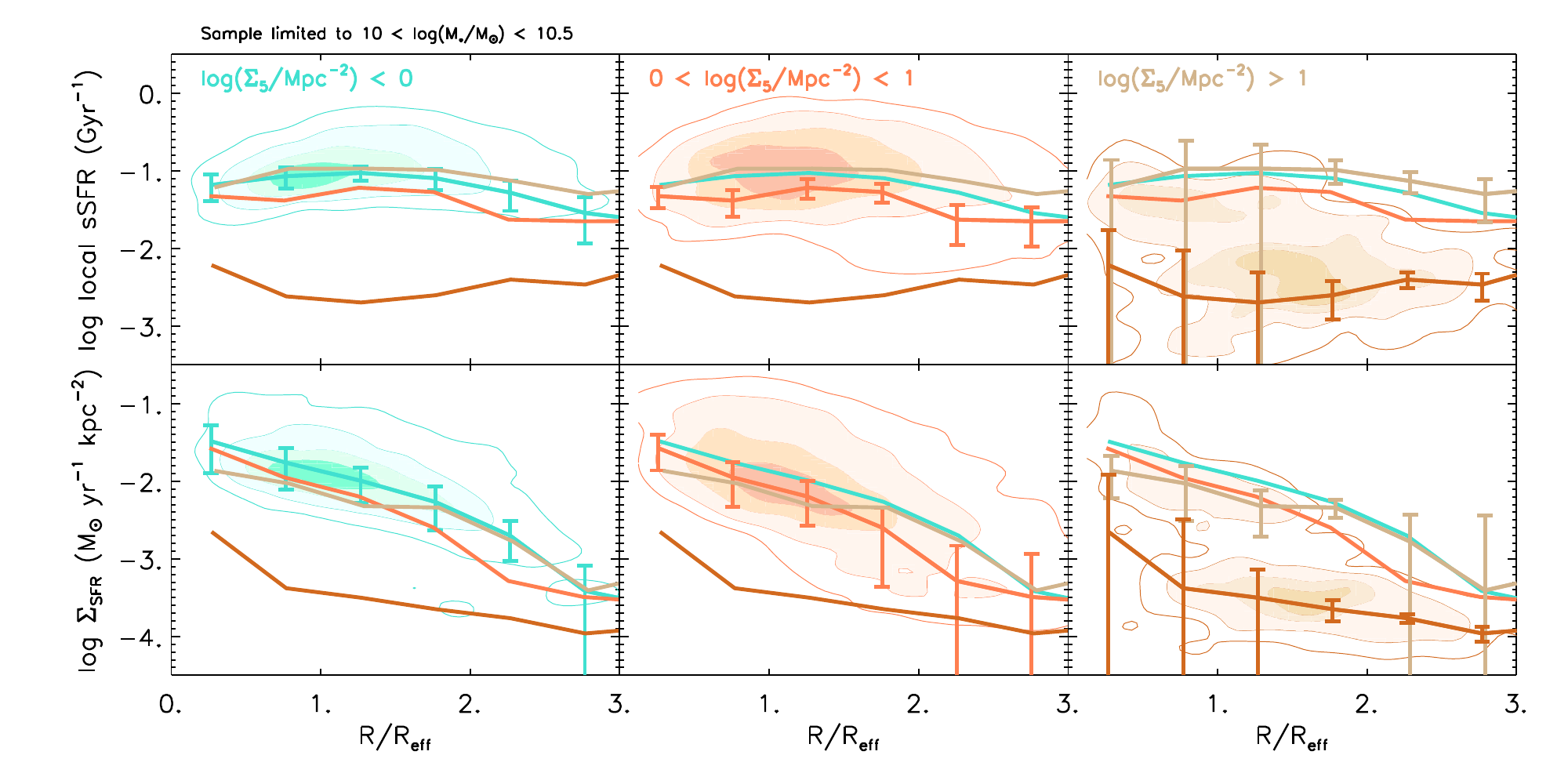}
\caption{
Top row: Radial profiles of specific star formation rate per spaxel in units of Gyr$^{-1}$ for SAMI galaxies with 10$<$log(M$_{*}$/M$_{\sun}$)$<$10.5 divided into three bins of environmental density according to \citet{Brough13}.  Each panel left-to-right shows radial profiles of all environmental density bins and contours showing the distribution of a single bin.  Each radial bin shows the median-level sSFR across galaxies in that sample; error bars are the standard error on the median, 
$1.253 \frac{\sigma_{\Sigma_{\text{SFR},n}}}{\sqrt{n}}$, 
where $\sigma_{\Sigma_{\text{SFR},n}}$ is the standard deviation of star formation rate surface densities for $n$ galaxies in a bin).
Bottom row: Radial profiles of star formation rate surface density in units of M$_{\sun}$ yr$^{-1}$ kpc$^{-2}$ for the same samples as top row.  Galaxies in denser environments systematically show lower sSFRs; the decreased sSFR of galaxies in denser environments is a direct result of lower SFRs.
For clarity, we split the high-density bin into two sets of galaxies: those contributing to the lower cloud in Figure~\ref{localSFMS_env} (and shown in its right-most panel) and those that mainly fall on the SFMS.  Median profiles for each set are shown in the panels, but only the distributions of the ``lower cloud'' galaxies are shown in the right-most panels here.  For galaxies that are experiencing some quenching (i.e. the lower cloud, dark brown profiles), the sSFR and SFR surface density profiles differ from isolated galaxies most significantly above 1.5 R$_{\text{eff}}$.
} 
 \label{localSSFR_env}
\end{figure*}

With the first subset of SAMI galaxies, \citet{Schaefer16} found that galaxies in denser environments show steeper SFR gradients, exactly what one would expect if interactions or ram-pressure stripping \citep{Gunn72} were to preferentially heat or remove gas from the outskirts of galaxies \citep[as shown by][]{Koopmann04a,Koopmann04b,Cortese12}.  Our sample of SAMI galaxies confirms this trend (see Figure~\ref{localSSFR_env}), again controlling for stellar mass by limiting to 10$<$log(M$_{*}$/M$_{\sun}$)$<$10.5, showing a set of galaxies in dense environments that are experiencing lower sSFR and $\Sigma_{\text{SFR}}$ at radii above 1.5 R$_{\text{eff}}$ compared to galaxies in low density bins.  The possible projection effects are too small to account for this difference.  Note that not all galaxies in our high density bin demonstrate this quenching behaviour, but our current sample is too small to compare the radial star-forming profiles of further subpopulations.
Schaefer et al. (submitted) further examined these quenching signatures relative to a variety of environmental indicators and found the dynamical mass of the parent halo to have the strongest effect, in line with studies of gas stripping such as \citet{Brown17}.

Galaxies in denser environments may also experience less gas accretion or may have gas in their halos heated by the intracluster medium \citep[e.g.][]{Larson80}, limiting further star formation due to a lack of cold gas.  If these strangulation-type mechanisms were dominating in our sample, we'd expect star formation rates to decrease everywhere, or perhaps more strikingly in the nuclei, where depletion times can be shorter \citep{Leroy13}.  Because star formation rates are most affected on the outskirts of the galaxy, we favor stripping as the dominant mechanism; however, we cannot rule out the possibility that these processes operate in tandem.

\citet{Bloom17} found that SAMI galaxies with kinematic asymmetries have more centrally-concentrated star formation, which may simply be due to galaxy-galaxy interactions triggering enhanced central star formation \citep[e.g.][]{Bekki11, Ellison13, Moreno15} and/or stripping gas from the outskirts.  Denser environments involve an increased likelihood of interactions, so it is unsurprising that we see a similar trend here.  However, the galaxies in dense environments seen in Figures~\ref{localSFMS_env} and~\ref{localSSFR_env} have decreased $\Sigma_{\text{SFR}}$, not increased, suggesting that outside-in quenching may be more likely than inside-out rejuvenation.

\subsection{Mass Quenching}
\label{massquenching}

\subsubsection{Quenching Effects of Galaxy Stellar Mass M$_{*}$}

In the previous two subsections, we eliminate any mass effect in order to study the variation in sSFR and SFR radial profiles across morphological types and environmental densities for galaxies with 10$<$log(M$_{*}$/M$_{\sun}$)$<$10.5.  Here we show the opposite, varying total stellar mass and limiting our sample to a single morphological classification and environmental densities 0$<$log($\Sigma_{5}$/Mpc$^{-2}$)$<$1 (orange in Figure~\ref{localSFMS_env}).  Galaxies may be more likely to quench at high stellar masses because they have massive halos; if a halo is massive enough ($\sim$10$^{12}$ M$_{\sun}$), infalling gas can be shock-heated to the virial temperature and is therefore inaccessible as fuel for star formation \citep{Birnboim03,Keres05,Dekel06,Woo13}.  Accordingly, observations have shown that decreased SFR in galaxies with high stellar masses is accompanied by a decrease in the cold gas reservoir \citep[e.g.][]{Tacconi13,Genzel15,Saintonge12,Saintonge16}.  Note that, in the absence of a direct measure of galaxy halo mass, we use total stellar mass as a proxy in this and following discussions; doing so should introduce $\lesssim$0.2 dex of scatter into the halo masses
\citep{Gu16}.

\begin{figure*}
\centering
\includegraphics[scale=0.9,trim=0.8cm 0.3cm 0cm 0.8cm]{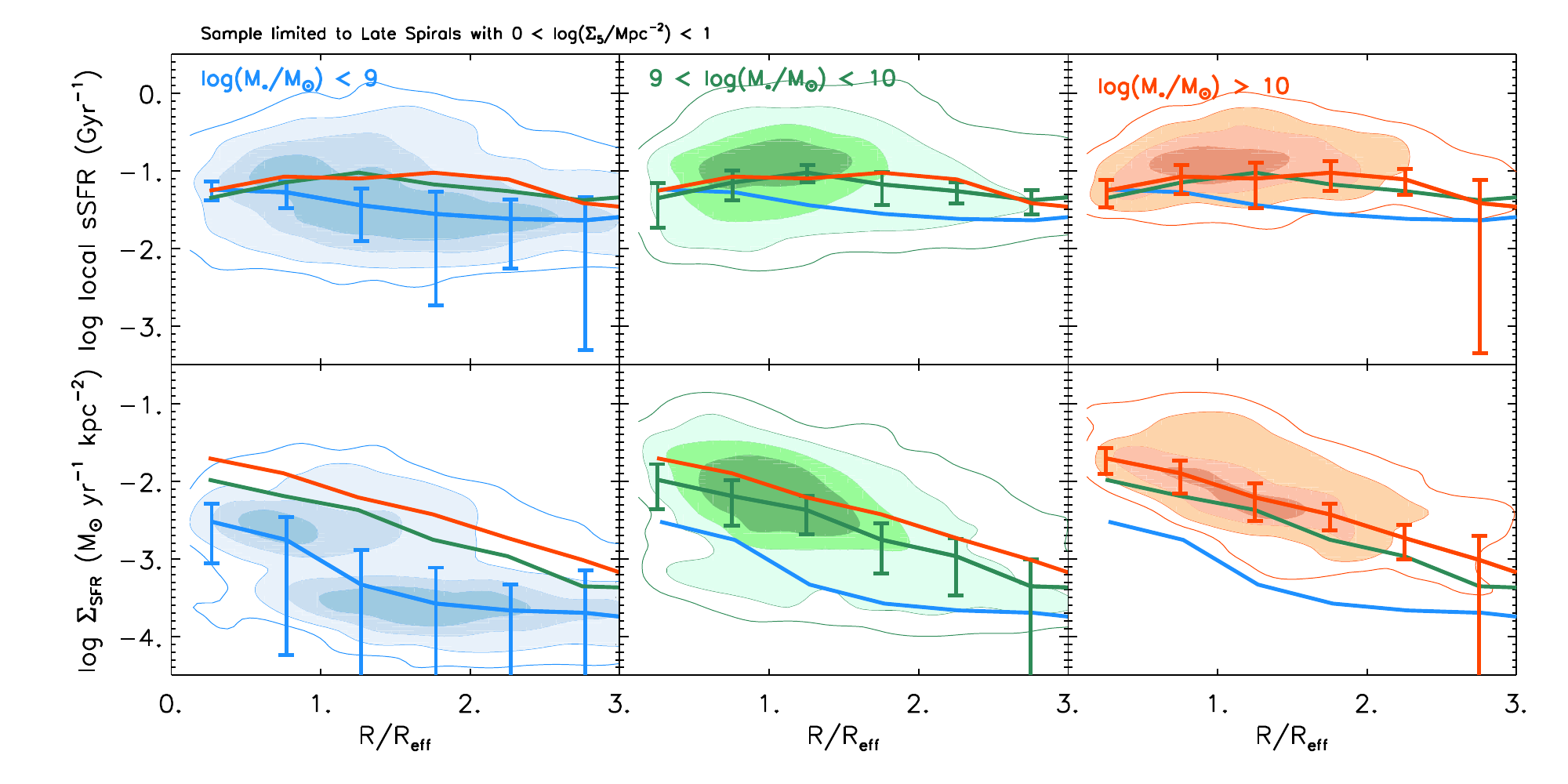}
\caption{
Top row: Radial profiles of specific star formation rate per spaxel in units of Gyr$^{-1}$ for SAMI galaxies classified as late-type spirals in environmental densities 0$<$log($\Sigma_{5}$/Mpc$^{-2})<$1 divided into three bins of stellar mass.  Each panel left-to-right shows radial profiles of all mass bins and contours showing the distribution of a single bin.  Each radial bin shows the median-level sSFR across galaxies in that sample; error bars are the standard error on the median, 
$1.253 \frac{\sigma_{\Sigma_{\text{SFR},n}}}{\sqrt{n}}$, 
where $\sigma_{\Sigma_{\text{SFR},n}}$ is the standard deviation of star formation rate surface densities for $n$ galaxies in a bin).
sSFR shows no correlation with stellar mass.  
Bottom row: Radial profiles of star formation rate surface density in units of M$_{\sun}$ yr$^{-1}$ kpc$^{-2}$ for the same samples as top row.  As expected, the $\Sigma_{\text{SFR}}$ values are higher at all radii for higher mass galaxies.  
} 
 \label{localSFMS_LS}
\end{figure*}

We investigate two particular morphological classifications.  Figure~\ref{localSFMS_LS} presents three mass bins of late-type spirals (purple in Figure~\ref{localSSFR_morph}), which show no significant variation between mass bins in the radial sSFR profiles.  The corresponding $\Sigma_{\text{SFR}}$ profiles (bottom row, Figure~\ref{localSFMS_LS}) show the expected correlation with mass across the entire radial distribution.  \citet{Schaefer16} also found no dependence on sSFR with stellar mass for star-forming galaxies (and that there is an increase in quiescent fraction at higher masses, as we saw in Section~\ref{global}).  The fact that we see no change in sSFR in star-forming galaxies with mass here suggests that if halo quenching occurs in late-type spirals, it must act at masses higher than we probe here.  When controlling for morphology, we see no evidence for the flattening in the SFMS seen by others \citep[e.g.][]{Karim11,Bauer13,Whitaker14,Lee15}.

Interestingly, galaxies in the lowest mass bin show similar star-forming behaviour to galaxies in dense environments (as seen in Figure~\ref{localSSFR_env}), even though this section limits the sample to the middle density bin.  Because of their low stellar masses, these galaxies may be more susceptible to stripping in moderate environments or might be satellites in a more massive halo.

\begin{figure*}
\centering
\includegraphics[scale=0.9,trim=0.8cm 0.3cm 0cm 0.8cm]{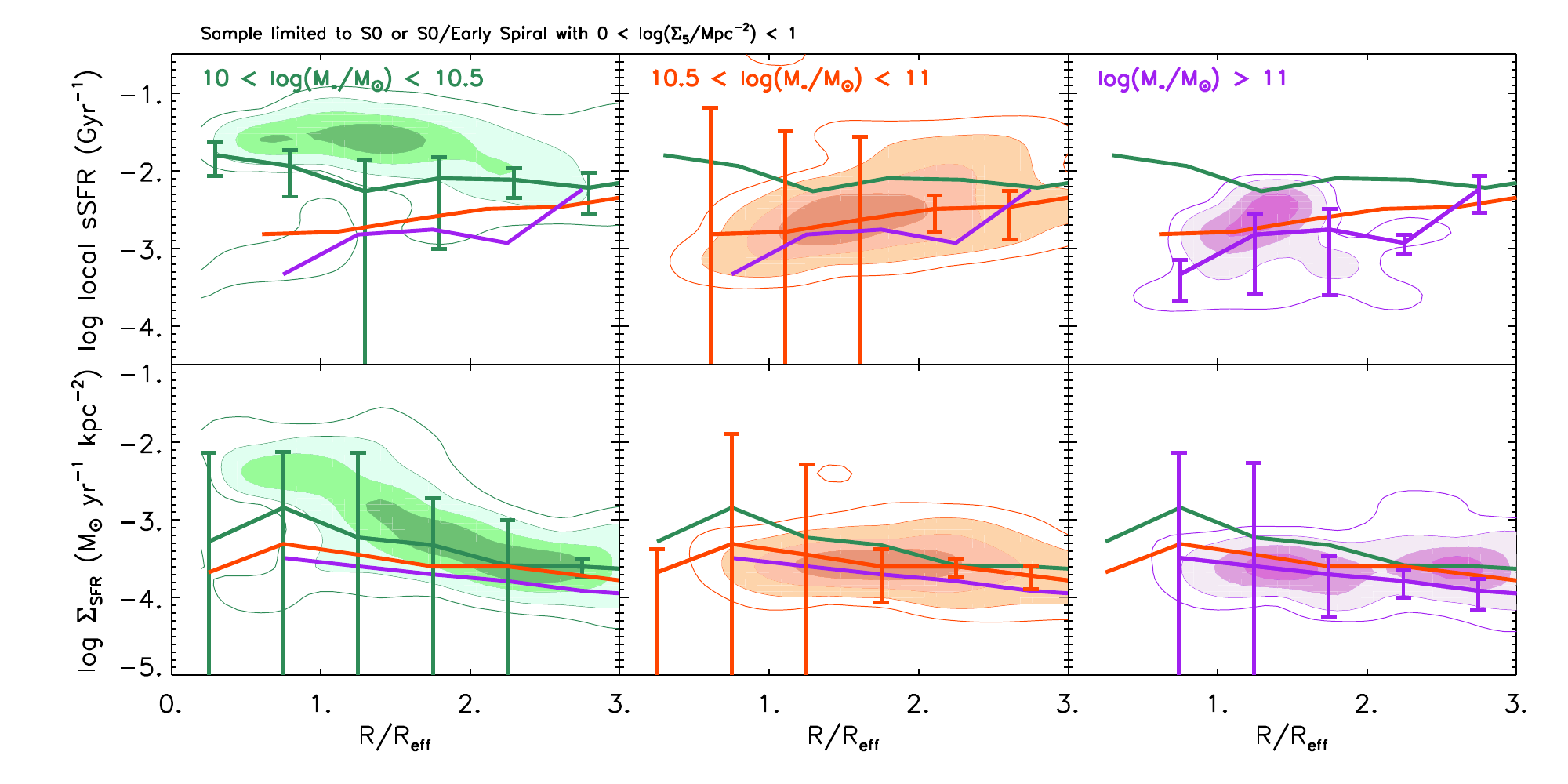}
\caption{
\textbf{Top row:} Radial profiles of specific star formation rate per spaxel in units of Gyr$^{-1}$ for SAMI galaxies classified as S0 or S0/early spirals in environmental densities 0$<$log($\Sigma_{5}$/Mpc$^{-2})<$1 divided into three bins of stellar mass.  Each panel left-to-right shows radial profiles of all mass bins and contours showing the distribution of a single bin.  Each radial bin shows the median-level sSFR across galaxies in that sample; error bars are the standard error on the median, 
$1.253 \frac{\sigma_{\Sigma_{\text{SFR},n}}}{\sqrt{n}}$, 
where $\sigma_{\Sigma_{\text{SFR},n}}$ is the standard deviation of star formation rate surface densities for $n$ galaxies in a bin).
Although not significant with this sample size, the sSFR may be decreasing with increasing galaxy mass, showing an increased likelihood of quenching.  
\textbf{Bottom row:} Radial profiles of star formation rate surface density in units of M$_{\sun}$ yr$^{-1}$ kpc$^{-2}$ for the same samples as top row.  
} 
 \label{localSFMS_SO}
\end{figure*}

To probe whether halo quenching might act at higher masses or in other morphological types, we also show radial sSFR and $\Sigma_{\text{SFR}}$ profiles of S0/early-type spirals (orange in Figure~\ref{localSSFR_morph}) in three mass bins (Figure~\ref{localSFMS_SO}).  Although this sample is too small to make firm conclusions, the decreasing sSFR levels with increasing stellar mass suggest that halo quenching could be occurring.  We do not have the sample here to determine whether halo quenching signatures might appear only at higher masses than our late-type spirals probe or only in galaxies with bulges.  

We note that in this section, we are using stellar mass as a proxy for halo mass.  By limiting the environmental densities, we avoid clusters, but may still include satellite galaxies in a more massive halo.  Of course, ``halo quenching'' is a phenomenon that could apply to isolated/central galaxies, satellite galaxies, or cluster members; we only test the first scenario here.

\subsubsection{Local Stellar Mass Surface Density Effects}

One proposed mechanism for quenching star formation is the local stellar mass surface density: as stellar mass builds up, it could potentially 
stabilize the nearby gas against star formation through 
excess turbulence from moderate gravitational collapse \citep{Martig09}, or through the increased velocity shear associated with a bulge's stellar density profile \citep{Federrath16}.  Indeed, recent observations have suggested that the central stellar mass surface density may be a stronger predictor of quenching than the total stellar mass \citep{Brinchmann04, Franx08, Bezanson09, Whitaker17}.  

If local stellar mass surface density is the key mechanism to suppress local star formation, we might expect that spaxels falling below the resolved star-forming main sequence would exhibit the highest stellar mass surface densities.  Figure~\ref{splitpops_localSFMS} shows no evidence for such a trend.  Quenched galaxies have a range of stellar mass surface densities.  Even amongst the main sequence population, spatial regions showing a depressed $\Sigma_{\text{SFR}}$ are as common or more at the lowest $\Sigma_{M*}$ values than the highest.  Further, the highest $\Sigma_{M*}$ spaxels are not merely missing because their star formation is below our detection limit: the distributions of $\Sigma_{M*}$ for spaxels with detected and undetected SFR are statistically indistinct.

\section{Discussion - What Causes Quenching?}
\label{discussion}

We have examined the effects of morphology, environmental density, and host galaxy stellar mass on SFR and sSFR profiles.  Our environmental results are the most straightforward (Section~\ref{environment}).  Dense environments host some galaxies with markedly lower SFRs and sSFRs at radii outside of 1.5 R$_{\text{eff}}$ (Figure~\ref{localSSFR_env}), compared to galaxies of similar stellar masses in less-dense environments.
Galaxies being quenched environmentally are therefore likely experiencing a physical process that limits star formation from the outside in, such as gas being stripped from the outer regions by the group or cluster halo, or from the increased likelihood of interactions.  

Our investigation of host galaxy stellar mass and local stellar mass surface density (Section~\ref{massquenching}) shows that these quantities are--on their own--insufficient to cause a galaxy to quench.  In Section~\ref{morph}, we show that early-type galaxies split into normal star-forming and ``quenched'' populations, demonstrating that the buildup of a bulge--again, on its own--is insufficient to trigger quenching.  Here we look at the differences between the two populations and the resulting star-forming behaviour of galaxies $>$3$\sigma$ below the global main sequence to gain insight into the causes of quenching.

The quenched population of early-type galaxies shows decreased sSFR and $\Sigma_{\text{SFR}}$ most strongly in the nuclear regions (Figure~\ref{localSSFR_morph}); analogous findings in the past have led to the term ``inside-out quenching'' and suggest star formation is likely to shut off first in the nuclei and progress outwards.  Two general processes could cause inside-out quenching:
\begin{itemize}
\item Nuclear gas reservoirs are depleted.  In this scenario, gas resupply might be cut off to the entire galaxy (i.e. through halo quenching), but higher star formation would lead to faster depletion in the nucleus.  Alternatively, gas from the nuclear regions might be preferentially evacuated through AGN or starburst-driven feedback.
\item Galaxy nuclei contain gas reservoirs that are stable against star formation.  In this scenario, feedback or the velocity shear associated with a steep potential well / high central stellar mass surface density might drive turbulence that prevents or slows the gravitational collapse of gas into stars.
\end{itemize}
Previous studies have seen evidence for both processes, but our results show that neither halo quenching nor bulge formation are independently capable of producing the quenched population.

If halo quenching were the root mechanism, our quenched population should have systematically higher stellar masses or environmental densities than the unquenched population, but the two samples show statistically similar distributions (Figure~\ref{splitpops_histograms}).  The two populations also show statistically similar distributions of central stellar mass surface density, ruling out the stellar bulge as the main driver of quenching.  If we include the galaxies below the SAMI \ha~detection limit (i.e. that appear in Section~\ref{global} but not Section~\ref{local}) as part of the quenched population, there are still no significant differences in distributions from the normal main sequence population.  Further, we see no evidence of a single ``cutoff'' stellar mass dictating whether an early-type galaxy stays on the main sequence or drops below.

\citet{Fang13} conclude that a high central stellar mass surface density is necessary but not sufficient for quenching, because they see some star-forming galaxies with high central densities, but no quenched galaxies with low central densities.  They propose a two-step quenching process, wherein gas accretion is halted when the galactic halo is hot enough to shock-heat the surrounding gas and internal gas is stabilized against star formation (via e.g. heating or turbulence), used up, or removed through another mechanism like AGN feedback; true quenching would only proceed when both mechanisms are active.  To test this scenario, we compare the total stellar mass and central stellar mass surface density of early-type galaxies on and below the main sequence in Figure~\ref{splitpops_massdistributions}, following Figure 5 of \citet{Fang13}.  We do not see evidence for the quenched galaxies residing at the highest central stellar mass surface densities for a given stellar mass, when limiting the sample to only early-type galaxies.  Note that we have removed the stellar mass limit from Section~\ref{splitpops} to include all early-type galaxies above/below the line 3$\sigma$ below the global main sequence in Figure~\ref{globalSFMS_env}.

\begin{figure}
\includegraphics[scale=0.92,trim=.5cm 0.8cm 0cm 0cm]{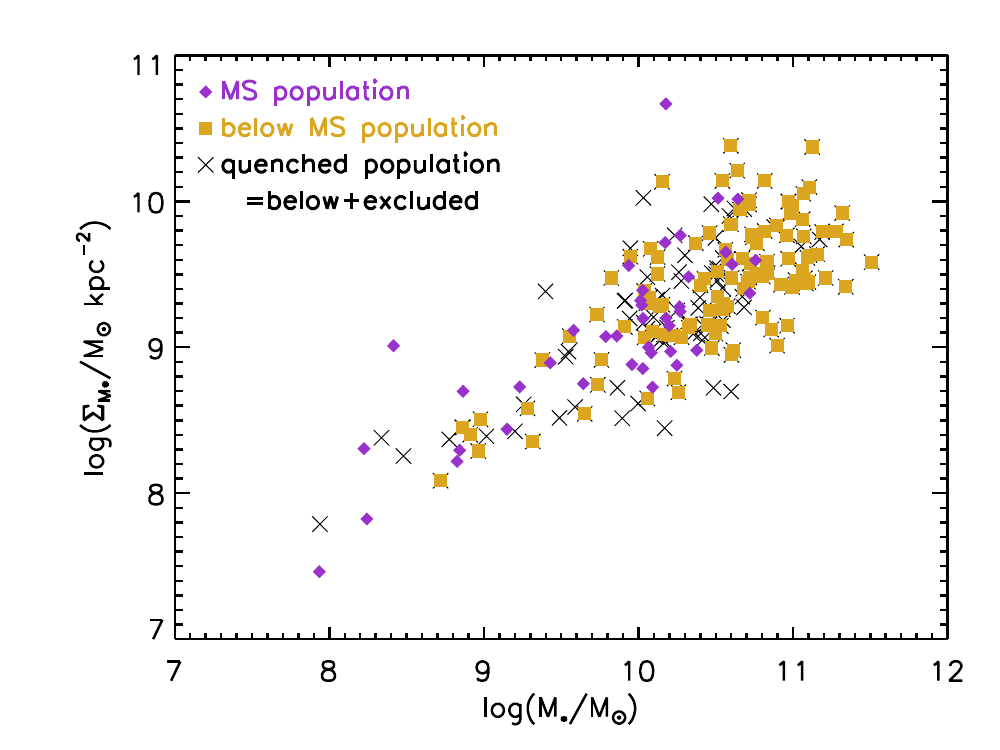}
\caption{
Central stellar mass surface densities versus total stellar masses of upper (purple diamonds), lower (gold squares), and lower+excluded/undetected galaxies (black crosses) from Section~\ref{splitpops}.  Our quenched galaxies do not have higher central stellar mass surface densities for at a given stellar mass. } 
\label{splitpops_massdistributions}
\end{figure}

Our results support a complex quenching process wherein more than one condition must be met, but the proposed two-step mechanism of \citet{Fang13} does not fully explain our split populations.  Halo quenching, if active, should be stifling the accretion for all galaxies in Figure~\ref{earlytype_splitpops} because there's no difference in galaxy mass between the quenched and unquenched populations.  If the supply of new gas has already been cut off and current star formation is gradually depleting the remaining fuel, there should be no dichotomy of $\Sigma_{\text{SFR}}$ profiles but a continuous spectrum.  If remaining gas is instead stabilized against star formation by the buildup of stellar mass, we would expect the quenched population to have higher stellar mass surface densities than the unquenched population.
Perhaps the critical mass at which halo quenching occurs has a large enough scatter that we are seeing merely a larger probability of being quenched at higher stellar masses.  We further propose that these bulges are not massive enough to stabilize a well-fueled star-forming disc against star formation, but as the fuel is depleted after halo quenching begins, the velocity shear suppresses star formation before it would otherwise gradually fade away.  

\section{Summary}
\label{summary} 

We describe the extinction maps, star formation masks, and clean star formation rate maps derived from \ha~for the SAMI Galaxy Survey, which are available for download through the SAMI Database at \url{datacentral.aao.gov.au} as part of the first SAMI Public Data Release.  We use these star formation rate maps, along with stellar mass maps created following \citet{Taylor11}, to examine the global and resolved star-forming main sequences of SAMI galaxies as a function of morphology \citep[as classified by][]{Cortese16}, environmental density \citep[according to][]{Brough13}, and host galaxy stellar mass.  We find:
\begin{itemize}
\item Below a stellar mass of $10^{10}$ M$_{\sun}$, only galaxies in medium to dense environments (log($\Sigma_{5}$/Mpc$^{-2}$) $>$0.5) fall below the star-forming main sequence.  Above this mass cutoff, isolated galaxies are also sometimes quenched.

\item Further below the main sequence, more galaxies show flat SFR surface density profiles.  This flattening is limited or uncommon within 1-3$\sigma$~of the main sequence, and does not appear common in our sample until $\sim$3$\sigma$~below the main sequence or more.  Galaxies lying just below the global main sequence 
are not experiencing a limited version of the same quenching mechanism that completely shuts off star formation in other galaxies.

\item Early-type galaxies split into two populations with different star formation behaviours.  Galaxies on the global star-forming main sequence show similar sSFR and $\Sigma_{\text{SFR}}$ profiles to late-type galaxies and lie just to the right of the late-type resolved star-forming main sequence.  Star formation in this population is not being quenched, but appears reduced only because the stellar mass in the bulge reduces the \textit{specific} SFRs.  Galaxies $>$3$\sigma$ below the global star-forming main sequence instead show significantly reduced sSFR and $\Sigma_{\text{SFR}}$ profiles, particularly in the nuclei, a distinct signature of inside-out quenching.  

\item The two early-type populations have statistically similar stellar masses and central stellar mass surface densities, showing that neither halo quenching nor bulge formation are independently sufficient to produce the split.  Our data favor a complex quenching approach similar to that proposed by \citet{Fang13}, wherein both halo mass and the local stellar mass surface densities play a role.  In our scenario, a galaxy grows until its massive halo shuts off further accretion; remaining gas continues to form stars until the velocity shear induced by the bulge is sufficient to suppress star formation.

\item Expanding on the sample of \citet{Schaefer16}, we confirm that galaxies in denser environments have overall lower sSFRs, and that this decrease is most pronounced in the outskirts of those galaxies.  These data support the scenario in which denser environments heat or strip gas on the outer edges of galaxies through close interactions or pressure from the intragroup/intracluster medium.

\end{itemize}

  
\section*{Acknowledgements}
We thank the referee for their thoughtful suggestions to improve the manuscript.
Support for AMM is provided by NASA through Hubble Fellowship grant \#HST-HF2-51377 awarded by the Space Telescope Science Institute, which is operated by the Association of Universities for Research in Astronomy, Inc., for NASA, under contract NAS5-26555.  
SMC acknowledges the support of an Australian Research Council Future Fellowship (FT100100457).
SB acknowledges funding support from the Australian Research Council through a Future Fellowship (FT140101166)
BC is the recipient of an Australian Research Council Future Fellowship (FT120100660).  
CF gratefully acknowledges funding provided by the Australian Research Council's Discovery Projects (grants DP150104329 and DP170100603).
MSO acknowledges the funding support from the Australian Research Council through a Future Fellowship Fellowship (FT140100255).
RMcD is the recipient of an Australian Research Council Future Fellowship (project number FT150100333).
NS acknowledges support of a University of Sydney Postdoctoral Research Fellowship.
JvdS is funded under Bland-Hawthorn's ARC Laureate Fellowship (FL140100278).

The SAMI Galaxy Survey is based on observations made at the Anglo-Australian Telescope. The Sydney-AAO Multi-object Integral field spectrograph (SAMI) was developed jointly by the University of Sydney and the Australian Astronomical Observatory. The SAMI input catalogue is based on data taken from the Sloan Digital Sky Survey, the GAMA Survey and the VST ATLAS Survey. The SAMI Galaxy Survey is funded by the Australian Research Council Centre of Excellence for All-sky Astrophysics (CAASTRO), through project number CE110001020, and other participating institutions. The SAMI Galaxy Survey website is http://sami-survey.org/ .

GAMA is a joint European-Australasian project based around a spectroscopic campaign using the Anglo-Australian Telescope. The GAMA input catalogue is based on data taken from the Sloan Digital Sky Survey and the UKIRT Infrared Deep Sky Survey. Complementary imaging of the GAMA regions is being obtained by a number of independent survey programmes including GALEX MIS, VST KiDS, VISTA VIKING, WISE, Herschel-ATLAS, GMRT and ASKAP providing UV to radio coverage. GAMA is funded by the STFC (UK), the ARC (Australia), the AAO, and the participating institutions. The GAMA website is http://www.gama-survey.org/.  This work is based on observations made with ESO Telescopes at the La Silla Paranal Observatory under programme IDs 179.A-2004 and 177.A-3016.

\bibliographystyle{mn2e}

\begin{thebibliography}{140}
\providecommand{\natexlab}[1]{#1}

\bibitem[{{Ahn} et~al.(2012)}]{Ahn12}
{Ahn} C.~P. et~al., 2012, \apjs, 203, 21

\bibitem[{{Allen} et~al.(2015)}]{Allen15}
{Allen} J.~T. et~al., 2015, \mnras, 446, 1567

\bibitem[{{Baldry} et~al.(2004){Baldry}, {Glazebrook}, {Brinkmann},
  {Ivezi{\'c}}, {Lupton}, {Nichol} \& {Szalay}}]{Baldry04}
{Baldry} I.~K., {Glazebrook} K., {Brinkmann} J., {Ivezi{\'c}} {\v Z}., {Lupton}
  R.~H., {Nichol} R.~C., {Szalay} A.~S., 2004, \apj, 600, 681

\bibitem[{{Baldry} et~al.(2012)}]{Baldry12}
{Baldry} I.~K. et~al., 2012, \mnras, 421, 621

\bibitem[{{Baldry} et~al.(2017)}]{Baldry17}
{Baldry} I.~K. et~al., 2017, ArXiv:1711.09139

\bibitem[{{Baldwin} et~al.(1981){Baldwin}, {Phillips} \& {Terlevich}}]{BPT}
{Baldwin} J.~A., {Phillips} M.~M., {Terlevich} R., 1981, \pasp, 93, 5

\bibitem[{{Balogh} \& {Morris}(2000)}]{Balogh00b}
{Balogh} M.~L., {Morris} S.~L., 2000, \mnras, 318, 703

\bibitem[{{Balogh} et~al.(1999){Balogh}, {Morris}, {Yee}, {Carlberg} \&
  {Ellingson}}]{Balogh99}
{Balogh} M.~L., {Morris} S.~L., {Yee} H.~K.~C., {Carlberg} R.~G., {Ellingson}
  E., 1999, \apj, 527, 54

\bibitem[{{Balogh} et~al.(2000){Balogh}, {Navarro} \& {Morris}}]{Balogh00a}
{Balogh} M.~L., {Navarro} J.~F., {Morris} S.~L., 2000, \apj, 540, 113

\bibitem[{{Bassett} et~al.(2017)}]{Bassett17}
{Bassett} R. et~al., 2017, \mnras, 470, 1991

\bibitem[{{Bauer} et~al.(2013)}]{Bauer13}
{Bauer} A.~E. et~al., 2013, \mnras, 434, 209

\bibitem[{{Bekki} \& {Couch}(2011)}]{Bekki11}
{Bekki} K., {Couch} W.~J., 2011, \mnras, 415, 1783

\bibitem[{{Belfiore} et~al.(2017{\natexlab{a}})}]{Belfiore17b}
{Belfiore} F. et~al., 2017{\natexlab{a}}, ArXiv e-prints

\bibitem[{{Belfiore} et~al.(2017{\natexlab{b}})}]{Belfiore17a}
{Belfiore} F. et~al., 2017{\natexlab{b}}, \mnras, 466, 2570

\bibitem[{{Bezanson} et~al.(2009){Bezanson}, {van Dokkum}, {Tal}, {Marchesini},
  {Kriek}, {Franx} \& {Coppi}}]{Bezanson09}
{Bezanson} R., {van Dokkum} P.~G., {Tal} T., {Marchesini} D., {Kriek} M.,
  {Franx} M., {Coppi} P., 2009, \apj, 697, 1290

\bibitem[{{Binette} et~al.(1994){Binette}, {Magris}, {Stasi{\'n}ska} \&
  {Bruzual}}]{Binette94}
{Binette} L., {Magris} C.~G., {Stasi{\'n}ska} G., {Bruzual} A.~G., 1994, \aap,
  292, 13

\bibitem[{{Birnboim} \& {Dekel}(2003)}]{Birnboim03}
{Birnboim} Y., {Dekel} A., 2003, \mnras, 345, 349

\bibitem[{{Bland-Hawthorn} et~al.(1991){Bland-Hawthorn}, {Sokolowski} \&
  {Cecil}}]{JBH91}
{Bland-Hawthorn} J., {Sokolowski} J., {Cecil} G., 1991, \pasp, 103, 906

\bibitem[{{Bland-Hawthorn} et~al.(2011)}]{BlandHawthorn11}
{Bland-Hawthorn} J. et~al., 2011, Optics Express, 19, 2649

\bibitem[{{Bloom} et~al.(2017)}]{Bloom17}
{Bloom} J.~V. et~al., 2017, \mnras, 465, 123

\bibitem[{{Brammer} et~al.(2009)}]{Brammer09}
{Brammer} G.~B. et~al., 2009, \apjl, 706, L173

\bibitem[{{Brinchmann} et~al.(2004){Brinchmann}, {Charlot}, {White},
  {Tremonti}, {Kauffmann}, {Heckman} \& {Brinkmann}}]{Brinchmann04}
{Brinchmann} J., {Charlot} S., {White} S.~D.~M., {Tremonti} C., {Kauffmann} G.,
  {Heckman} T., {Brinkmann} J., 2004, \mnras, 351, 1151

\bibitem[{{Brough} et~al.(2013)}]{Brough13}
{Brough} S. et~al., 2013, \mnras, 435, 2903

\bibitem[{{Brown} et~al.(2017)}]{Brown17}
{Brown} T. et~al., 2017, \mnras, 466, 1275

\bibitem[{{Bruzual} \& {Charlot}(2003)}]{BruzualCharlot03}
{Bruzual} G., {Charlot} S., 2003, \mnras, 344, 1000

\bibitem[{{Bryant} et~al.(2011){Bryant}, {O'Byrne}, {Bland-Hawthorn} \&
  {Leon-Saval}}]{Bryant11}
{Bryant} J.~J., {O'Byrne} J.~W., {Bland-Hawthorn} J., {Leon-Saval} S.~G., 2011,
  \mnras, 415, 2173

\bibitem[{{Bryant} et~al.(2014){Bryant}, {Bland-Hawthorn}, {Fogarty},
  {Lawrence} \& {Croom}}]{Bryant14}
{Bryant} J.~J., {Bland-Hawthorn} J., {Fogarty} L.~M.~R., {Lawrence} J.~S.,
  {Croom} S.~M., 2014, \mnras, 438, 869

\bibitem[{{Bryant} et~al.(2015)}]{Bryant15}
{Bryant} J.~J. et~al., 2015, \mnras, 447, 2857

\bibitem[{{Bundy} et~al.(2015)}]{MaNGA}
{Bundy} K. et~al., 2015, \apj, 798, 7

\bibitem[{{Calzetti}(2001)}]{Calzetti01}
{Calzetti} D., 2001, \pasp, 113, 1449

\bibitem[{{Calzetti} et~al.(2000){Calzetti}, {Armus}, {Bohlin}, {Kinney},
  {Koornneef} \& {Storchi-Bergmann}}]{Calzetti00}
{Calzetti} D., {Armus} L., {Bohlin} R.~C., {Kinney} A.~L., {Koornneef} J.,
  {Storchi-Bergmann} T., 2000, \apj, 533, 682

\bibitem[{{Cano-D{\'{\i}}az} et~al.(2016)}]{Cano-Diaz16}
{Cano-D{\'{\i}}az} M. et~al., 2016, \apjl, 821, L26

\bibitem[{{Cardelli} et~al.(1989){Cardelli}, {Clayton} \&
  {Mathis}}]{Cardelli89}
{Cardelli} J.~A., {Clayton} G.~C., {Mathis} J.~S., 1989, \apj, 345, 245

\bibitem[{{Carroll} et~al.(1992){Carroll}, {Press} \& {Turner}}]{Carroll92}
{Carroll} S.~M., {Press} W.~H., {Turner} E.~L., 1992, \araa, 30, 499

\bibitem[{{Catal{\'a}n-Torrecilla} et~al.(2015)}]{Catalan-Torrecilla15}
{Catal{\'a}n-Torrecilla} C. et~al., 2015, \aap, 584, A87

\bibitem[{{Cervi{\~n}o} \& {Luridiana}(2004)}]{Cervino04}
{Cervi{\~n}o} M., {Luridiana} V., 2004, \aap, 413, 145

\bibitem[{{Cervi{\~n}o} \& {Valls-Gabaud}(2003)}]{Cervino03}
{Cervi{\~n}o} M., {Valls-Gabaud} D., 2003, \mnras, 338, 481

\bibitem[{{Chabrier}(2003)}]{Chabrier03}
{Chabrier} G., 2003, \pasp, 115, 763

\bibitem[{{Cortese} et~al.(2012)}]{Cortese12}
{Cortese} L. et~al., 2012, \aap, 544, A101

\bibitem[{{Cortese} et~al.(2014)}]{Cortese14}
{Cortese} L. et~al., 2014, \apjl, 795, L37

\bibitem[{{Cortese} et~al.(2016)}]{Cortese16}
{Cortese} L. et~al., 2016, \mnras, 463, 170

\bibitem[{{Croom} et~al.(2012)}]{Croom12}
{Croom} S.~M. et~al., 2012, \mnras, 421, 872

\bibitem[{{da Cunha} et~al.(2008){da Cunha}, {Charlot} \& {Elbaz}}]{daCunha08}
{da Cunha} E., {Charlot} S., {Elbaz} D., 2008, \mnras, 388, 1595

\bibitem[{{da Silva} et~al.(2014){da Silva}, {Fumagalli} \&
  {Krumholz}}]{daSilva14}
{da Silva} R.~L., {Fumagalli} M., {Krumholz} M.~R., 2014, \mnras, 444, 3275

\bibitem[{{Daddi} et~al.(2007)}]{Daddi07}
{Daddi} E. et~al., 2007, \apj, 670, 156

\bibitem[{{Daddi} et~al.(2010)}]{Daddi10b}
{Daddi} E. et~al., 2010, \apjl, 714, L118

\bibitem[{{Davies} et~al.(2016)}]{Davies16}
{Davies} L.~J.~M. et~al., 2016, \mnras, 461, 458

\bibitem[{{de Jong} et~al.(2015)}]{deJong15}
{de Jong} J.~T.~A. et~al., 2015, \aap, 582, A62

\bibitem[{{Dekel} \& {Birnboim}(2006)}]{Dekel06}
{Dekel} A., {Birnboim} Y., 2006, \mnras, 368, 2

\bibitem[{{Dopita}(1976)}]{Dopita76}
{Dopita} M.~A., 1976, \apj, 209, 395

\bibitem[{{Driver}(2017)}]{Driver17}
{Driver} S., 2017, \mnras, submitted

\bibitem[{{Driver} et~al.(2011)}]{GAMA}
{Driver} S.~P. et~al., 2011, \mnras, 413, 971

\bibitem[{{Edge} et~al.(2013){Edge}, {Sutherland}, {Kuijken}, {Driver},
  {McMahon}, {Eales} \& {Emerson}}]{Edge13}
{Edge} A., {Sutherland} W., {Kuijken} K., {Driver} S., {McMahon} R., {Eales}
  S., {Emerson} J.~P., 2013, The Messenger, 154, 32

\bibitem[{{Edge} et~al.(2014){Edge}, {Sutherland} \& {VIKING Team}}]{Edge14}
{Edge} A., {Sutherland} W., {VIKING Team}, 2014, VizieR Online Data Catalog,
  2329

\bibitem[{{Elbaz} et~al.(2007)}]{Elbaz07}
{Elbaz} D. et~al., 2007, \aap, 468, 33

\bibitem[{{Ellison} et~al.(2011){Ellison}, {Nair}, {Patton}, {Scudder},
  {Mendel} \& {Simard}}]{Ellison11}
{Ellison} S.~L., {Nair} P., {Patton} D.~R., {Scudder} J.~M., {Mendel} J.~T.,
  {Simard} L., 2011, \mnras, 416, 2182

\bibitem[{{Ellison} et~al.(2013){Ellison}, {Mendel}, {Patton} \&
  {Scudder}}]{Ellison13}
{Ellison} S.~L., {Mendel} J.~T., {Patton} D.~R., {Scudder} J.~M., 2013, \mnras,
  435, 3627

\bibitem[{{Fabian}(2012)}]{Fabian12}
{Fabian} A.~C., 2012, \araa, 50, 455

\bibitem[{{Fang} et~al.(2013){Fang}, {Faber}, {Koo} \& {Dekel}}]{Fang13}
{Fang} J.~J., {Faber} S.~M., {Koo} D.~C., {Dekel} A., 2013, \apj, 776, 63

\bibitem[{{Federrath}(2015)}]{Federrath15}
{Federrath} C., 2015, \mnras, 450, 4035

\bibitem[{{Federrath} et~al.(2016)}]{Federrath16}
{Federrath} C. et~al., 2016, \apj, 832, 143

\bibitem[{{Federrath} et~al.(2017)}]{Federrath17}
{Federrath} C. et~al., 2017, \mnras, 468, 3965

\bibitem[{{Feldmann}(2017)}]{Feldmann17}
{Feldmann} R., 2017, \mnras, 470, L59

\bibitem[{{Franx} et~al.(2008){Franx}, {van Dokkum}, {F{\"o}rster Schreiber},
  {Wuyts}, {Labb{\'e}} \& {Toft}}]{Franx08}
{Franx} M., {van Dokkum} P.~G., {F{\"o}rster Schreiber} N.~M., {Wuyts} S.,
  {Labb{\'e}} I., {Toft} S., 2008, \apj, 688, 770-788

\bibitem[{{Garc{\'{\i}}a-Benito} et~al.(2015)}]{CALIFA_DR2}
{Garc{\'{\i}}a-Benito} R. et~al., 2015, \aap, 576, A135

\bibitem[{{Geha} et~al.(2012){Geha}, {Blanton}, {Yan} \& {Tinker}}]{Geha12}
{Geha} M., {Blanton} M.~R., {Yan} R., {Tinker} J.~L., 2012, \apj, 757, 85

\bibitem[{{Genzel} et~al.(2015)}]{Genzel15}
{Genzel} R. et~al., 2015, \apj, 800, 20

\bibitem[{{Goddard} et~al.(2017{\natexlab{a}})}]{Goddard17b}
{Goddard} D. et~al., 2017{\natexlab{a}}, \mnras, 466, 4731

\bibitem[{{Goddard} et~al.(2017{\natexlab{b}})}]{Goddard17a}
{Goddard} D. et~al., 2017{\natexlab{b}}, \mnras, 465, 688

\bibitem[{{Gonz{\'a}lez Delgado} et~al.(2016)}]{GonzalezDelgado16}
{Gonz{\'a}lez Delgado} R.~M. et~al., 2016, \aap, 590, A44

\bibitem[{{Green} et~al.(2014)}]{Green14}
{Green} A.~W. et~al., 2014, \mnras, 437, 1070

\bibitem[{{Green} et~al.(2017)}]{Green17}
{Green} A.~W. et~al., 2017, ArXiv:1707.08402

\bibitem[{{Groves} et~al.(2012){Groves}, {Brinchmann} \& {Walcher}}]{Groves12}
{Groves} B., {Brinchmann} J., {Walcher} C.~J., 2012, \mnras, 419, 1402

\bibitem[{{Gu} et~al.(2016){Gu}, {Conroy} \& {Behroozi}}]{Gu16}
{Gu} M., {Conroy} C., {Behroozi} P., 2016, \apj, 833, 2

\bibitem[{{Gunawardhana} et~al.(2013)}]{Gunawardhana13}
{Gunawardhana} M.~L.~P. et~al., 2013, \mnras, 433, 2764

\bibitem[{{Gunn} \& {Gott}(1972)}]{Gunn72}
{Gunn} J.~E., {Gott} III J.~R., 1972, \apj, 176, 1

\bibitem[{{Haines} et~al.(2008){Haines}, {Gargiulo} \& {Merluzzi}}]{Haines08}
{Haines} C.~P., {Gargiulo} A., {Merluzzi} P., 2008, \mnras, 385, 1201

\bibitem[{{Hampton} et~al.(2017)}]{Hampton17}
{Hampton} E.~J. et~al., 2017, \mnras, 470, 3395

\bibitem[{{Hinshaw} et~al.(2009)}]{Hinshaw09}
{Hinshaw} G. et~al., 2009, \apjs, 180, 225

\bibitem[{{Ho} et~al.(2016)}]{LZIFU}
{Ho} I.~T. et~al., 2016, \apss, 361, 280

\bibitem[{{Ibarra-Medel} et~al.(2016)}]{IbarraMedel16}
{Ibarra-Medel} H.~J. et~al., 2016, \mnras, 463, 2799

\bibitem[{{Karim} et~al.(2011)}]{Karim11}
{Karim} A. et~al., 2011, \apj, 730, 61

\bibitem[{{Kelvin} et~al.(2014)}]{Kelvin14}
{Kelvin} L.~S. et~al., 2014, \mnras, 439, 1245

\bibitem[{{Kennicutt} et~al.(1994){Kennicutt}, {Tamblyn} \&
  {Congdon}}]{Kennicutt94}
{Kennicutt} Jr. R.~C., {Tamblyn} P., {Congdon} C.~E., 1994, \apj, 435, 22

\bibitem[{{Kere{\v s}} et~al.(2005){Kere{\v s}}, {Katz}, {Weinberg} \&
  {Dav{\'e}}}]{Keres05}
{Kere{\v s}} D., {Katz} N., {Weinberg} D.~H., {Dav{\'e}} R., 2005, \mnras, 363,
  2

\bibitem[{{Kewley} et~al.(2006){Kewley}, {Groves}, {Kauffmann} \&
  {Heckman}}]{Kewley06}
{Kewley} L.~J., {Groves} B., {Kauffmann} G., {Heckman} T., 2006, \mnras, 372,
  961

\bibitem[{{Koopmann} \& {Kenney}(2004{\natexlab{a}})}]{Koopmann04b}
{Koopmann} R.~A., {Kenney} J.~D.~P., 2004{\natexlab{a}}, \apj, 613, 866

\bibitem[{{Koopmann} \& {Kenney}(2004{\natexlab{b}})}]{Koopmann04a}
{Koopmann} R.~A., {Kenney} J.~D.~P., 2004{\natexlab{b}}, \apj, 613, 851

\bibitem[{{Krumholz} et~al.(2017){Krumholz}, {Kruijssen} \&
  {Crocker}}]{Krumholz17}
{Krumholz} M.~R., {Kruijssen} J.~M.~D., {Crocker} R.~M., 2017, \mnras, 466,
  1213

\bibitem[{{Larson} et~al.(1980){Larson}, {Tinsley} \& {Caldwell}}]{Larson80}
{Larson} R.~B., {Tinsley} B.~M., {Caldwell} C.~N., 1980, \apj, 237, 692

\bibitem[{{Lee} et~al.(2015)}]{Lee15}
{Lee} N. et~al., 2015, \apj, 801, 80

\bibitem[{{Leroy} et~al.(2013)}]{Leroy13}
{Leroy} A.~K. et~al., 2013, \aj, 146, 19

\bibitem[{{Li} et~al.(2015)}]{Li15}
{Li} C. et~al., 2015, \apj, 804, 125

\bibitem[{{Mapelli}(2015)}]{Mapelli15}
{Mapelli} M., 2015, Galaxies, 3, 192

\bibitem[{{Martig} et~al.(2009){Martig}, {Bournaud}, {Teyssier} \&
  {Dekel}}]{Martig09}
{Martig} M., {Bournaud} F., {Teyssier} R., {Dekel} A., 2009, \apj, 707, 250

\bibitem[{{Mo} et~al.(1998){Mo}, {Mao} \& {White}}]{Mo98}
{Mo} H.~J., {Mao} S., {White} S.~D.~M., 1998, \mnras, 295, 319

\bibitem[{{Moreno} et~al.(2015){Moreno}, {Torrey}, {Ellison}, {Patton},
  {Bluck}, {Bansal} \& {Hernquist}}]{Moreno15}
{Moreno} J., {Torrey} P., {Ellison} S.~L., {Patton} D.~R., {Bluck} A.~F.~L.,
  {Bansal} G., {Hernquist} L., 2015, \mnras, 448, 1107

\bibitem[{{Mu{\~n}oz-Mateos} et~al.(2007){Mu{\~n}oz-Mateos}, {Gil de Paz},
  {Boissier}, {Zamorano}, {Jarrett}, {Gallego} \& {Madore}}]{MunozMateos07}
{Mu{\~n}oz-Mateos} J.~C., {Gil de Paz} A., {Boissier} S., {Zamorano} J.,
  {Jarrett} T., {Gallego} J., {Madore} B.~F., 2007, \apj, 658, 1006

\bibitem[{{Muldrew} et~al.(2012)}]{Muldrew12}
{Muldrew} S.~I. et~al., 2012, \mnras, 419, 2670

\bibitem[{{Nelson} et~al.(2016)}]{Nelson16}
{Nelson} E.~J. et~al., 2016, \apj, 828, 27

\bibitem[{{Noeske} et~al.(2007{\natexlab{a}})}]{Noeske07b}
{Noeske} K.~G. et~al., 2007{\natexlab{a}}, \apjl, 660, L47

\bibitem[{{Noeske} et~al.(2007{\natexlab{b}})}]{Noeske07a}
{Noeske} K.~G. et~al., 2007{\natexlab{b}}, \apjl, 660, L43

\bibitem[{{Owers} et~al.(2017)}]{Owers17}
{Owers} M.~S. et~al., 2017, \mnras, 468, 1824

\bibitem[{{Peng} et~al.(2015){Peng}, {Maiolino} \& {Cochrane}}]{Peng15}
{Peng} Y., {Maiolino} R., {Cochrane} R., 2015, \nat, 521, 192

\bibitem[{{Peng} et~al.(2012){Peng}, {Lilly}, {Renzini} \& {Carollo}}]{Peng12}
{Peng} Y.~j., {Lilly} S.~J., {Renzini} A., {Carollo} M., 2012, \apj, 757, 4

\bibitem[{{Peng} et~al.(2010)}]{Peng10}
{Peng} Y.~j. et~al., 2010, \apj, 721, 193

\bibitem[{{P{\'e}rez} et~al.(2013)}]{Perez13}
{P{\'e}rez} E. et~al., 2013, \apjl, 764, L1

\bibitem[{{Renzini} \& {Peng}(2015)}]{RenziniPeng15}
{Renzini} A., {Peng} Y.~j., 2015, \apjl, 801, L29

\bibitem[{{Richards} et~al.(2016)}]{Richards16}
{Richards} S.~N. et~al., 2016, \mnras, 455, 2826

\bibitem[{{Saintonge} et~al.(2012)}]{Saintonge12}
{Saintonge} A. et~al., 2012, \apj, 758, 73

\bibitem[{{Saintonge} et~al.(2016)}]{Saintonge16}
{Saintonge} A. et~al., 2016, \mnras, 462, 1749

\bibitem[{{S{\'a}nchez} et~al.(2012)}]{CALIFA}
{S{\'a}nchez} S.~F. et~al., 2012, \aap, 538, A8

\bibitem[{{S{\'a}nchez} et~al.(2013)}]{Sanchez13}
{S{\'a}nchez} S.~F. et~al., 2013, \aap, 554, A58

\bibitem[{{Santini} et~al.(2009)}]{Santini09}
{Santini} P. et~al., 2009, \aap, 504, 751

\bibitem[{{Sarzi} et~al.(2008)}]{Sarzi08}
{Sarzi} M. et~al., 2008, in J.H. {Knapen}, T.J. {Mahoney}, A.~{Vazdekis}, eds,
  Pathways Through an Eclectic Universe. Astronomical Society of the Pacific
  Conference Series, Vol. 390, p. 218

\bibitem[{{Schaefer} et~al.(2017)}]{Schaefer16}
{Schaefer} A.~L. et~al., 2017, \mnras, 464, 121

\bibitem[{{Scoville} et~al.(2016)}]{Scoville16}
{Scoville} N. et~al., 2016, \apj, 820, 83

\bibitem[{{Sharp} et~al.(2006)}]{Sharp06}
{Sharp} R. et~al., 2006, in Society of Photo-Optical Instrumentation Engineers
  (SPIE) Conference Series. \procspie, Vol. 6269, p. 62690G

\bibitem[{{Sharp} et~al.(2015)}]{Sharp15}
{Sharp} R. et~al., 2015, \mnras, 446, 1551

\bibitem[{{Shull} \& {McKee}(1979)}]{Shull79}
{Shull} J.~M., {McKee} C.~F., 1979, \apj, 227, 131

\bibitem[{{Silverman} et~al.(2015)}]{Silverman15}
{Silverman} J.~D. et~al., 2015, \apjl, 812, L23

\bibitem[{{Speagle} et~al.(2014){Speagle}, {Steinhardt}, {Capak} \&
  {Silverman}}]{Speagle14}
{Speagle} J.~S., {Steinhardt} C.~L., {Capak} P.~L., {Silverman} J.~D., 2014,
  \apjs, 214, 15

\bibitem[{{Strickland}(2002)}]{Strickland02}
{Strickland} D., 2002, in R.~{Fusco-Femiano}, F.~{Matteucci}, eds, Chemical
  Enrichment of Intracluster and Intergalactic Medium. Astronomical Society of
  the Pacific Conference Series, Vol. 253, p. 387

\bibitem[{{Tacchella} et~al.(2015)}]{Tacchella15}
{Tacchella} S. et~al., 2015, Science, 348, 314

\bibitem[{{Tacchella} et~al.(2016{\natexlab{a}})}]{Tacchella16b}
{Tacchella} S., {Dekel} A., {Carollo} C.~M., {Ceverino} D., {DeGraf} C.,
  {Lapiner} S., {Mandelker} N., {Primack} J.~R., 2016{\natexlab{a}}, \mnras,
  458, 242

\bibitem[{{Tacchella} et~al.(2016{\natexlab{b}})}]{Tacchella16a}
{Tacchella} S., {Dekel} A., {Carollo} C.~M., {Ceverino} D., {DeGraf} C.,
  {Lapiner} S., {Mandelker} N., {Primack Joel} R., 2016{\natexlab{b}}, \mnras,
  457, 2790

\bibitem[{{Tacconi} et~al.(2010)}]{Tacconi10}
{Tacconi} L.~J. et~al., 2010, \nat, 463, 781

\bibitem[{{Tacconi} et~al.(2013)}]{Tacconi13}
{Tacconi} L.~J. et~al., 2013, \apj, 768, 74

\bibitem[{{Taylor} et~al.(2011)}]{Taylor11}
{Taylor} E.~N. et~al., 2011, \mnras, 418, 1587

\bibitem[{{Toomre}(1964)}]{Toomre64}
{Toomre} A., 1964, \apj, 139, 1217

\bibitem[{{Tully} et~al.(1982){Tully}, {Mould} \& {Aaronson}}]{Tully82}
{Tully} R.~B., {Mould} J.~R., {Aaronson} M., 1982, \apj, 257, 527

\bibitem[{{Vazdekis} et~al.(2010)}]{Vazdekis10}
{Vazdekis} A., {S{\'a}nchez-Bl{\'a}zquez} P., {Falc{\'o}n-Barroso} J.,
  {Cenarro} A.~J., {Beasley} M.~A., {Cardiel} N., {Gorgas} J., {Peletier}
  R.~F., 2010, \mnras, 404, 1639

\bibitem[{{Veilleux} \& {Osterbrock}(1987)}]{VO87}
{Veilleux} S., {Osterbrock} D.~E., 1987, \apjs, 63, 295

\bibitem[{{Whitaker} et~al.(2012){Whitaker}, {van Dokkum}, {Brammer} \&
  {Franx}}]{Whitaker12}
{Whitaker} K.~E., {van Dokkum} P.~G., {Brammer} G., {Franx} M., 2012, \apjl,
  754, L29

\bibitem[{{Whitaker} et~al.(2014)}]{Whitaker14}
{Whitaker} K.~E. et~al., 2014, \apj, 795, 104

\bibitem[{{Whitaker} et~al.(2017)}]{Whitaker17}
{Whitaker} K.~E. et~al., 2017, \apj, 838, 19

\bibitem[{{White} \& {Frenk}(1991)}]{White91}
{White} S.~D.~M., {Frenk} C.~S., 1991, \apj, 379, 52

\bibitem[{{Woo} et~al.(2013)}]{Woo13}
{Woo} J. et~al., 2013, \mnras, 428, 3306

\bibitem[{{Wuyts} et~al.(2013)}]{Wuyts13}
{Wuyts} S. et~al., 2013, \apj, 779, 135

\bibitem[{{Zhou} et~al.(2017)}]{Zhou17}
{Zhou} L. et~al., 2017, \mnras, 470, 4573

\end{thebibliography}


\appendix
\section{Comparing Global Star Formation Rates}
\label{globalsfcorr}

In Section~\ref{global}, we presented the star-forming main sequence of galaxies using global measures of star formation calculated by summing the SAMI star formation rate maps.  These star formation rate maps are designed to explore a clean sample of spatially-resolved star formation and mask out any spaxels with spectra that are not classified as `star-forming'.  This choice should render our global star formation rates as lower limits in cases where AGN or shocks contaminate some regions, or where star formation is present but weak.  

To confirm that the star formation masks are the dominant inaccuracy in our data, we compare our global star formation rates to those from the GAMA survey \citep{Gunawardhana13,Davies16,Driver17}.  We choose three representative SFR indicators from \citet{Davies16}: extinction-corrected \ha~luminosity, which is directly comparable to our method; MAGPHYS, which performs full SED-fitting on 21 bands of photometry from the ultraviolet to the far-infrared; and radiative transfer (RT), which uses dust-corrected NUV luminosities.  By comparing to these multiwavelength indicators, we can also confirm if \ha~emission does a reasonable job tracing star formation in our galaxies and if our dust corrections are sufficient.

\begin{figure*}
\centering
\includegraphics[scale=0.95,trim=0.8cm 0.3cm 0cm 0.8cm]{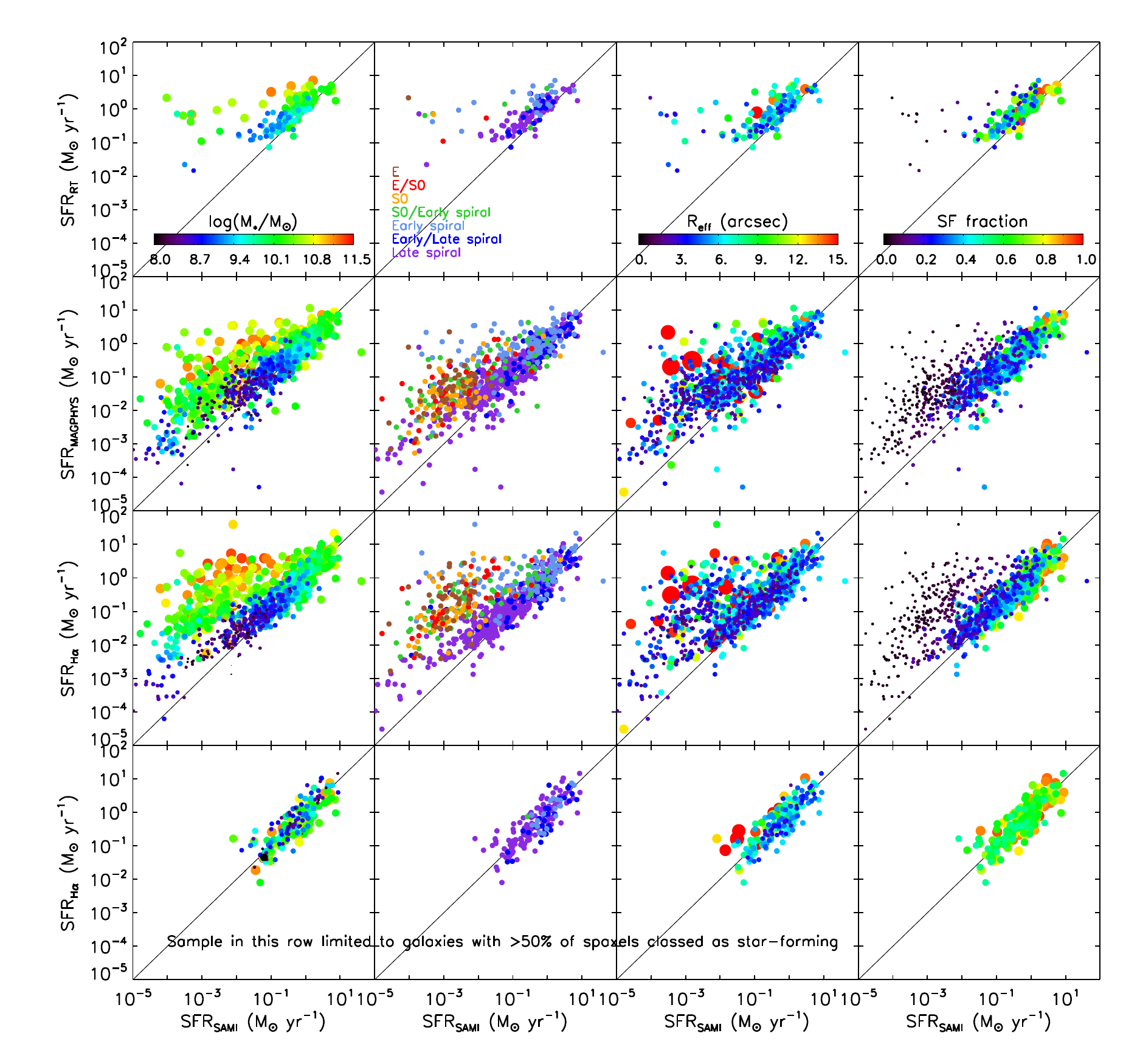}
\caption{
Comparisons of global star formation rates calculated from SAMI SFR maps (horizontal axes) to three methods of calculating global SFRs from GAMA data \citep{Gunawardhana13, Davies16}: the radiative transfer method (SFR$_{RT}$, top row), MAGPHYS \citep[SFR$_{MAGPHYS}$, second row;][]{daCunha08, Driver17}, and \ha~flux (SFR$_{H\alpha}$\_orig, third and fourth rows).  Note that all GAMA SFR measurements have been multiplied by 1.53 to account for the difference between a Chabrier initial mass function (IMF; in GAMA) and a Salpeter IMF (in this work).  The columns colour-code each galaxy by different properties to demonstrate what biases might be present in SAMI SFRs: (left to right) stellar mass, morphology, effective radius, and the fraction of spaxels in our map classified as star-forming and included in the global SFR calculation.  The right column clearly demonstrates that the most discrepant galaxies in our sample are those with low star-forming fractions.  The bottom row repeats the second row but limits the sample of galaxies to those with star-forming fractions greater than 50\%; this sample matches the GAMA SFRs well.  A small effect remains showing that the largest galaxies (R$_{\text{eff}}\sim15$\arcsec, large red points in third plot of bottom row) are still missing some star formation outside of the SAMI aperture.
} 
 \label{compareSFRs}
\end{figure*}

Figure~\ref{compareSFRs} compares these three different star formation rate measures from GAMA to our SAMI global SFRs, colour-coded by different galaxy properties.  Many SAMI galaxies have SFRs underestimated relative to the GAMA measures; the right column demonstrates that these galaxies are underestimated because most of the spaxels are not classified as star-forming in our star formation masks.  These spaxels may be excluded because of possible shock or AGN contamination (which could also affect the \ha-based measurements from GAMA) or because the signal-to-noise in the SAMI spectrum is too low to be confident that star formation dominates.  We emphasize that the latter is likely the case for our lowest SFR galaxies (early-type galaxies and low mass late-type galaxies); weak star formation may not produce detectable \ha~emission in a single spaxel.  The appropriate way to calculate the global SFRs for low-SNR galaxies is by summing all spaxels of the SAMI data cube into a single spectrum and then measuring \ha~emission, rather than measuring the \ha~flux in each spaxel and summing that.  Global properties from aperture-summed spectra, including SFRs, will be presented in a future catalog.  For now, we are confident that the SFRs presented here are otherwise correct and appropriate for a clean spatially-resolved analysis of star formation.

\end{document}